\documentclass[12pt]{article}
\usepackage{hyperref}
\usepackage{graphicx}
\usepackage{subcaption}
\usepackage{cite}
\usepackage{geometry}
\usepackage{url}
\usepackage{booktabs}
\usepackage{authblk} 

\title{Modeling Urban Air Quality Using Taxis as Sensors}

\author[1]{Anastasios Noulas}
\author[1]{Yasin Acikmese}
\author[5]{Charles QC LI}
\author[4]{Milan Y. Patel}
\author[2]{Shazia’Ayn Babul}
\author[4]{Ronald C. Cohen}
\author[2]{Renaud Lambiotte}
\author[3]{Marta C. Gonz\'{a}lez}

\affil[1]{The Alan Turing Institute, UK \\anoulas@turing.ac.uk}
\affil[1]{Firefly, USA\\yasin@fireflyon.com}
\affil[2]{Mathematical Institute, University of Oxford, UK\\renaud.lambiotte@maths.ox.ac.uk}
\affil[3]{Departments of City and Regional Planning and Civil and Environmental Engineering, University of California, Berkeley, USA\\martag@berkeley.edu}
\affil[4]{Department of Chemistry, University of California, Berkeley, USA\\rccohen@berkeley.edu}
\affil[5]{College of Environmental Design, University of California, Berkeley, USA\\liqianchuan9321@berkeley.edu}

\begin{document}

\maketitle

\begin{abstract}
Monitoring urban air quality with high spatiotemporal resolution continues to pose significant challenges. We investigate the use of taxi fleets as mobile sensing platforms, analyzing over 100 million PM2.5 readings from 3,000+ vehicles across six major U.S. cities during one year. Our findings show that taxis provide fine-grained, street-level air quality insights while ensuring city-wide coverage. We further explore urban air quality modeling using traffic congestion, built environment, and human mobility data to predict pollution variability. Our results highlight geography-specific seasonal patterns and demonstrate that models based solely on traffic and wind speeds effectively capture a city’s pollution dynamics. This study establishes taxi fleets as a scalable, near-real-time air quality monitoring tool, offering new opportunities for environmental research and data-driven policymaking.
\end{abstract}

\section{Introduction}
Poor environmental and atmospheric conditions in urban areas are among the leading causes of premature death and chronic respiratory conditions such as asthma and lung cancer~\cite{li2018function, landrigan2018pollution, cohen2017estimates}. This affects billions of people residing in urban and peri-urban environments worldwide~\cite{unesco2022urban}. A major concern for citizens, urban authorities, and policymakers is the ability to accurately and efficiently monitor urban air quality~\cite{who2021air}. One of the most widely adopted approaches to large-scale air quality monitoring involves citizen-led projects~\cite{castell2017can, egli2022citizenscience}, which have emerged to address limitations in scalability and maintenance costs associated with government-operated sensing infrastructure. These citizen initiatives have been inspired by the success of crowdsourcing efforts over the past two decades, which have enabled data collection across domains including mapping, urban social activity, and human mobility~\cite{haklay2013citizen}. 

Despite their critical value, these projects face significant challenges, including the need for sustained citizen participation, along with deployment and operational hurdles~\cite{maag2018sensebox}. In recent years, the concept of leveraging taxi fleets as urban sensing platforms has gained attention due to their advantages in scale, spatial resolution, and cost-effectiveness~\cite{o2019quantifying, hasenfratz2012participatory}. A growing body of work~\cite{judalet2021mapping, schuetz2023lowcost, liu2022spatiotemporal, novak2020utilizing, lee2021air} has shown that taxis and other vehicles can serve as effective platforms for air quality monitoring. Operating continuously across time and space within cities, taxis and ride-sharing vehicles offer a high degree of spatiotemporal coverage, enabling the collection of environmental data with fine granularity~\cite{aplin2021moving}. Unlike dedicated monitoring vehicles deployed by governments, research institutions, or private organizations~\cite{testi2024big}, taxis do not require explicit commissioning to traverse city streets—a process that adds substantial logistical and economic burden.

Here, we investigate the use of taxi fleets for urban air quality sensing through data collected by Firefly, a mobile advertising and technology platform. While there have been previous deployment scenarios that have utilized taxis as a sensing platform~\cite{sun2022high, wu2020application}, in this work we take a step further introducing a dataset of a much larger scale. 
We analyze over 100 million PM2.5 sensor readings collected from more than 3,000 vehicles operating in six major U.S. cities over a 12-month period (2022–2023). We demonstrate how existing vehicle fleets can be transformed into a large-scale air quality sensing network when supported by a sustainable business model~\cite{cheng2021scalable}. Using vehicle-based sensing, we capture the temporal variation in PM2.5 levels, including seasonal patterns and major environmental events~\cite{zhou2021urban}. Across the spatial dimension, we show how taxis can provide highly granular, street-level air quality measurements while maintaining broad citywide coverage~\cite{kamionka2020mobile}. We further investigate new possibilities in urban air quality modeling made feasible by large-scale mobile sensing. In particular, we assess the predictability of pollution levels based on urban indicators such as traffic congestion, built environment characteristics, and human mobility. Our analysis explores the trade-offs between localized prediction accuracy and the availability of supporting PM2.5 data~\cite{song2022airpred}. Visualizing PM2.5 concentrations at the city level, we identify distinct seasonal and geographic trends. We show that models using traffic congestion and wind speed data alone can effectively capture a city's characteristic pollution dynamics~\cite{zheng2013uair}.

Our findings suggest that taxi fleets can serve as operationally efficient, near-real-time air quality monitoring systems with the potential for global scalability.  
We expect these data to further advance research into the interplay between vehicular traffic, urban activity, and pollution levels~\cite{Miao2022, Shu2022, Sun2021, Wang2020, Wanke2020, Westmoreland2007, zhou2018impact}, and to support the development of mobility models and simulators~\cite{cai2021, wang2019, kim2020, santos2021, yuan2020, zhang2017mobility} that integrate pollution dispersion with urban dynamics. Moreover,
while the link between air quality and socioeconomic disparities~\cite{fecht2015, hajat2015, bravo2016, clark2014, hajat2013, bell2005environmental} are beyond the scope of this study, we hope these data can support future research addressing this urgent public health issue. 
\begin{figure}
  \centering
  \begin{subfigure}[htbp]{1.0\columnwidth}
    \includegraphics[width=\textwidth]{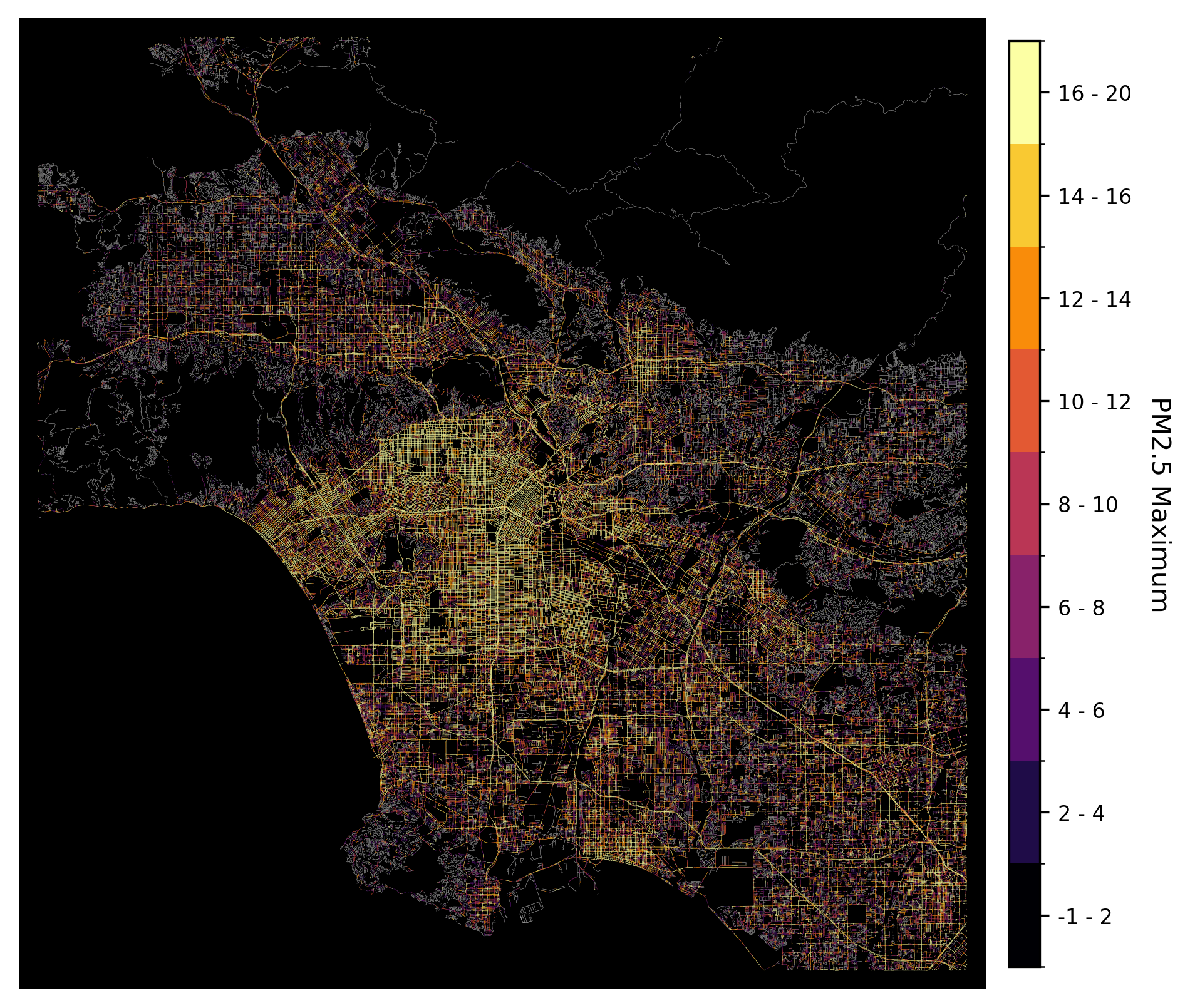}
    \label{fig:lanet}
  \end{subfigure}
  \caption{Urban Street Network visualization of the Los Angeles area. Color intensity is set according to the max PM2.5  record in each street segment. Areas of the network without available records have been colored in gray.}
  \label{fig:streetnets}
\end{figure}

\section{Results}
\subsection*{Instrumenting Taxi Fleets with PM2.5 sensors}
\label{sec:properties}
In recent years, Firefly\footnote{\url{www.fireflyon.com}} has developed and deployed a rapidly expanding technology platform integrating a digital display on top of taxis and ride-sharing vehicles with cloud-orchestrated, edge-based software system that enables real-time communication across a vehicular network. Although primarily designed to support an outdoor advertising business, the developed technology platform has allowed the deployment of environmental sensors to collect contextual signals from the vehicle’s surroundings as it moves through the city.
A key component of this system is a low-cost dust sensor embedded in the display (see S.I. Fig.~\ref{fig:sunrise}). The display features an Android board for data processing and communication purposes and is equipped with a set of sensors including a GPS sensor and a laser-based PM2.5 counter, which have been used to collect the data employed in the present work. The sensor measures PM2.5 concentrations of airborne particles with diameters of 2.5 micrometers or smaller, expressed in $\mu g/m^{3}$. Each sensor records PM2.5 levels approximately every 60 seconds, with a remotely adjustable sampling rate. As Firefly-equipped taxis traverse the city, covering extensive portions of the street network over time, their data, combined with GPS-based mobility information, provides a highly granular, dynamic view of urban air pollution. 
\begin{figure}[htbp]
  \centering
 \includegraphics[width=0.9\columnwidth]{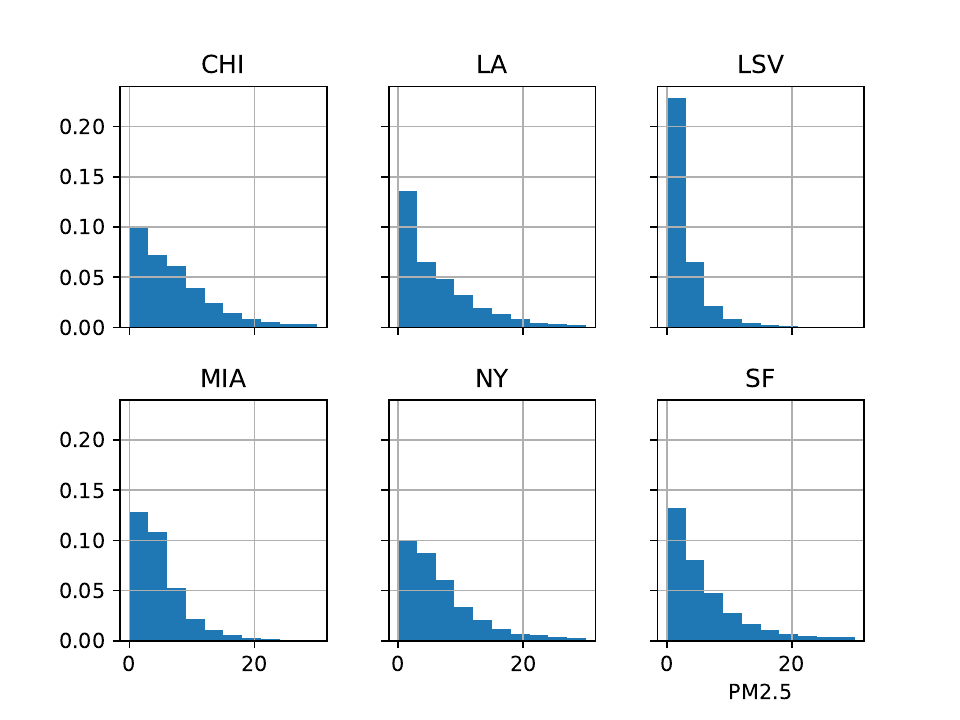}
  \caption{Per City PM2.5 Probability Density Functions  (0-30 value range).}
  \label{fig:cityhistograms}
\end{figure}
\subsection*{Data visualization and its basic properties}
In Fig.~\ref{fig:streetnets}, we visualize the dataset in the Greater Los Angeles area showing the maximum PM2.5 record during the time window under consideration (July 31st 2022 to August 1st 2023). Immediately, two key observations can be made. Firstly, the large heterogeneity of pollution levels emerging in the various locations of the city, highlighting how novel analytical insights can be obtained when viewing air quality patterns at high spatial resolution. Secondly, higher levels of pollution are recorded around dense urban cores, as well as the main arteries of the street network that form the traffic backbone of the city. 
\begin{figure}[htbp] 
  \centering
  \includegraphics[width=0.9\columnwidth]{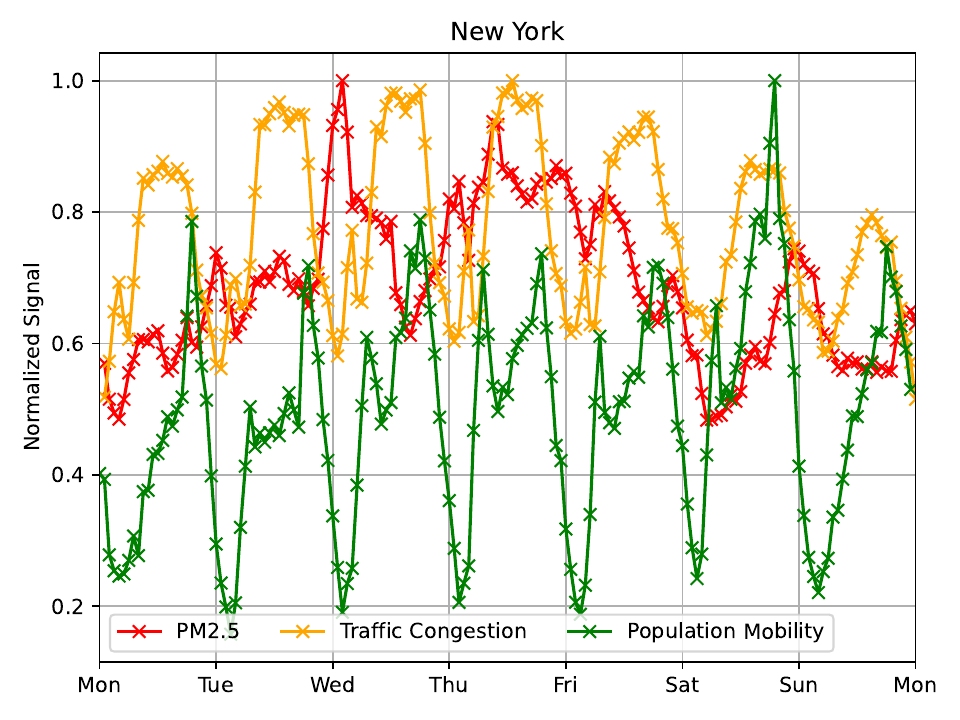}
  \caption{Aggregated weekly patterns of urban mobility and congestion signals plotted against PM2.5 levels in New York City. All data points are normalized according to the respective max value observed in the 168 dimensional vectors corresponding to the hours of a week.}
  \label{fig:weeklynyc}
\end{figure}
In Fig.~\ref{fig:cityhistograms}, we plot the probability distributions of PM2.5 readings in histogram form across the eight cities we study in the dataset (value range filtered to the 0-30 range covering $\approx 95\%$ of the data). While all cities follow a common pattern with the probability dropping significantly as PM2.5 values rise, there are also notable variations across cities 
with some having their probability mass shifted more towards higher values implying a wider spread of PM2.5 across the urban terrain.
Important questions arising in this setting revolve around the understanding of what urban structural as well as environmental and population activity characteristics raise the probability of higher pollution concentrations in a city.  Taxi-based measurements are inherently biased towards areas with higher vehicular and population mobility. However, they offer a significant advantage in spatial resolution terms. In contrast, government-installed monitoring stations are typically deployed across wider geographic regions, which can overlook PM2.5 variations within small regions of the same urban area (see S.I.~\ref{sec:supporting} for comparisons with EPA and Purple Air sensors). 
\begin{figure}[htbp] 
    \centering 

    \begin{subfigure}[b]{1.0\textwidth} 
        \centering 
        \includegraphics[width=\linewidth]{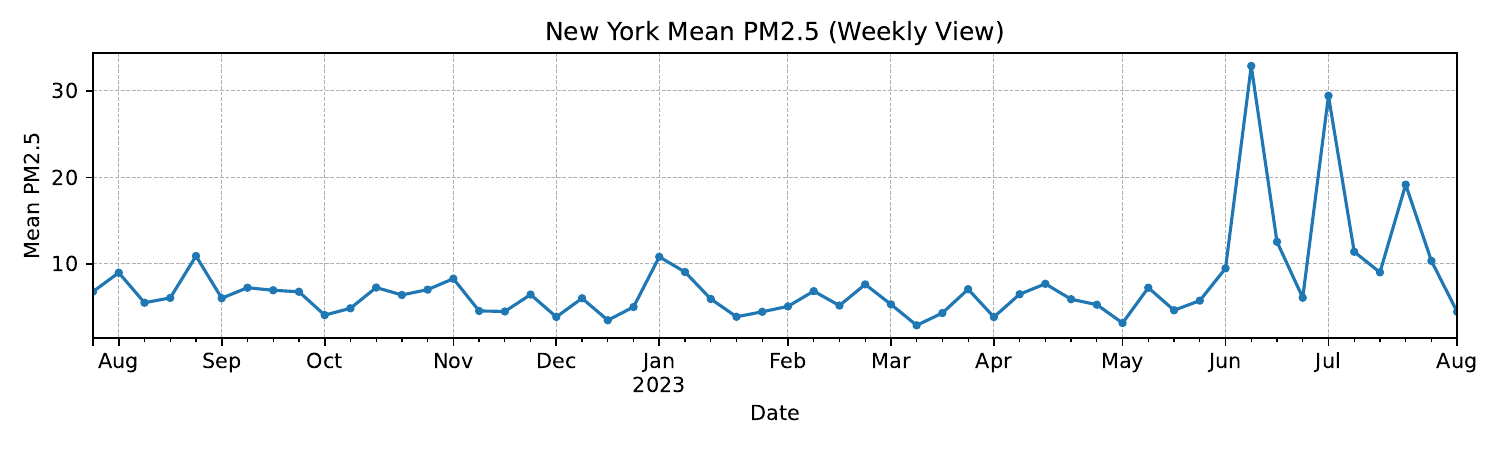} 
        \caption{Weekly Mean PM2.5 Levels} 
        \label{fig:nyc_weekly_mean} 
    \end{subfigure}
    \hfill 
    \begin{subfigure}[b]{1.0\textwidth} 
        \centering 
        \includegraphics[width=\linewidth]{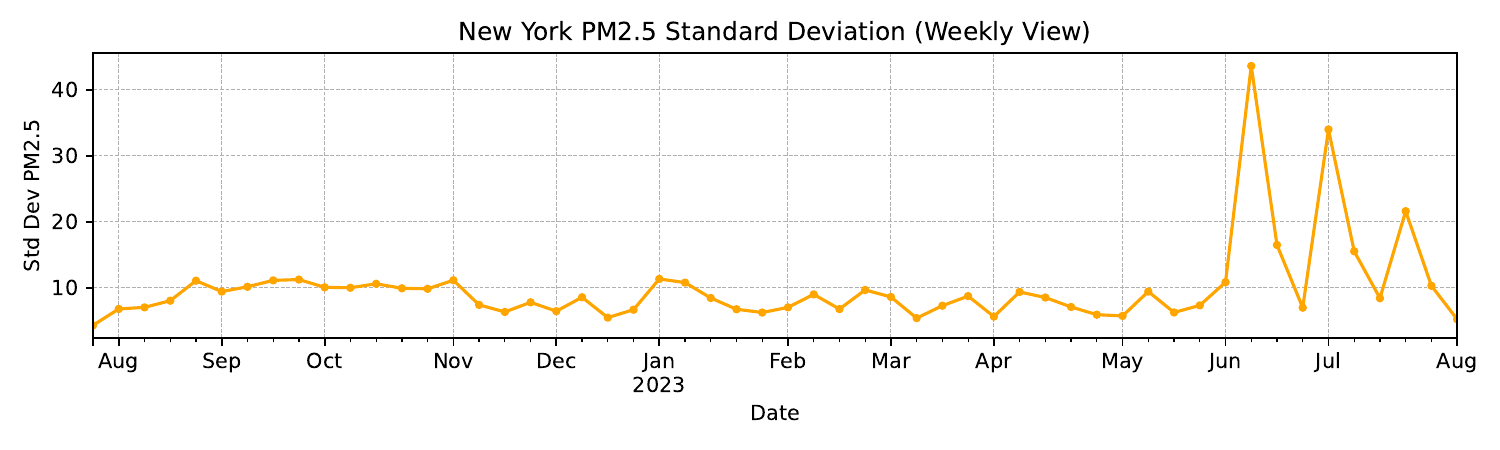} 
        \caption{Weekly PM2.5 Standard Deviation} 
        \label{fig:nyc_weekly_std} 
    \end{subfigure}

    \caption{Seasonal variations of PM2.5 levels in New York City. (a) Shows the weekly aggregated mean, while (b) shows the weekly aggregated standard deviation.}
    \label{fig:nyc_seasonal_combined} 

\end{figure}
Taking into account the temporal dimension, in Fig.~\ref{fig:weeklynyc}, we report the mean levels of PM2.5 recorded by Firefly taxis in New York City for each hour of the week, where each data point corresponds to the mean level of pollution observed at that hour of the week, aggregating over all weeks included in the span of the dataset. We normalize each data point with respect to the maximum observation during the 168-hour time window of a week and compare this signal with population mobility fluctuations and traffic congestion signal (see Section~{\ref{sec:modeling}} for formal definitions of all data sources). We observe daily periodicity patterns in PM2.5 levels with characteristic peaks and minimums. So is the case for population mobility and congestion, all of which follow characteristic temporal variations driven by the well-studied periodical nature of human activity patterns in the urban domain~\cite{jiang2016timegeo, kaltenbrunner2010urban, hasan2013spatiotemporal}. A natural question that can arise in this setting regards the relationship between the various temporal signals in the quest to recover potential synchronicities or even causal effects between human activity and PM2.5 levels. Doing so has proven challenging due to the sparsity of the data when these signals are disaggregated across large time windows. We will be focusing therefore on assessing the usefulness of congestion information and population flows in predicting PM2.5 at fine geographic scales. 
There are complex atmospheric physical and chemical phenomena taking place in the formation of PM2.5. A common misconception is that vehicular emissions are the direct source of fine particulate matter. However, a significant portion of PM2.5 is formed secondarily through atmospheric chemical reactions that lead to particle formation or growth~\cite{seinfeld2016atmospheric}. This contrasts with primary pollutants, such as nitrogen oxides (NOx) and black carbon, which are emitted directly from sources and exhibit the highest concentrations near their point of origin before dispersing into the background atmosphere~\cite{IPCC2021}. While some precursor gases contributing to secondary PM2.5 formation are indeed emitted by vehicles, studies in U.S. cities indicate that the majority of PM2.5 precursors originate from other sources, including biogenic emissions from vegetation (e.g., trees and shrubs) and various anthropogenic activities such as those from restaurants, fireplaces, and industrial processes~\cite{Zhang2015}. Given the timescale of several hours required for these secondary chemical reactions to occur, the spatial correlation between PM2.5 concentrations and their initial emission sources is often weak~\cite{FinlaysonPitts2000}. With these observations at hand, our aim in the following paragraphs is to combine signals that are proxies to the intensity of human urban activity, historic PM2.5 observations and descriptors of the built environment so as to assess the predictability of PM2.5 levels at various levels of spatial resolution.

A macroscopic temporal perspective on urban air quality can be obtained when observing longer time series and seasonal views of the data. Fig.~\ref{fig:nyc_weekly_mean} we present weekly mean PM2.5 levels in the New York region across the course of the year featuring an unusual increase in air pollution levels in early June 2023, a phenomenon known to have been induced by forest fires in Canada~\cite{thurston2023evaluation,wang2024severe, guardian_canada_wildfire_2023}. The reader can note that despite a short dip of recovery, increased levels of PM2.5 are sustained through the beginning of July when urban activity and fireworks have been reported to adversely influence air quality in urban areas~\cite{Fan2021}. In Fig.~\ref{fig:nyc_weekly_std}, we report the standard deviations observed during across weeks of the same time period, where a strong variance signal in the data is also noted.

From the perspective of seasonal variation, atmospheric aerosol can also be understood as a product of chemical formation, removal by deposition, and transport via winds~\cite{Pruppacher2010}. The rate of chemical aerosol production is likely to exhibit seasonal variations, potentially increasing during periods with higher solar irradiance~\cite{Carslaw2010}. Furthermore, the relative importance of different precursor emissions can vary throughout the year, with biogenic emissions potentially dominating during summer months and sources like residential fireplaces becoming more significant in winter~\cite{Kelly2012}. Both vertical atmospheric mixing and horizontal winds play critical roles in the dispersion and transport of aerosols. Strong vertical mixing leads to a rapid dilution of emissions, contrasting with conditions of weak mixing, such as those often experienced on cold winter days where localized pollution sources like fireplace emissions are more noticeable. For instance, the distinct smell of a neighbor's fireplace on a calm winter day exemplifies limited vertical dispersion~\cite{Stull1988}. Similarly, higher wind speeds facilitate the transport of emissions over greater distances, influencing the location of secondary aerosol formation. Conversely, aerosol measured at a specific location during periods of strong winds may have originated from sources located further away from the observation point~\cite{Stohl1998}.

In summary, diffusion of micro-particles and their concentration patterns in the atmospheric realm of a city may vary due to weather conditions (e.g. wind patterns, humidity and temperature) but also due to anthropogenic activity, namely industrial activity, vehicle traffic, as well as residential and other urban activities. Next we attempt to uncover these patterns in a modeling context considering two distinct scenarios each based on a different method of data aggregation. First, we aggregate data on a monthly basis to predict PM2.5 levels in a high spatial resolution environment within the city. Then, we aggregate data at the city level, to obtain denser representations across the temporal axis, aiming to recover diurnal and seasonal patterns of PM2.5 levels. In both scenarios we discuss the efficacy of various information signals in the prediction task as well as variations observed across different urban environments.

\subsection{Predicting PM2.5 across Fine Spatial Scales}
\label{sec:modeling}

Next, we present a predictive framework designed to estimate mean PM2.5 levels within a given area by leveraging a diverse array of signals characterizing the local urban environment and human activity. We employ en ensemble supervised learning method (Random Forests~\cite{breiman2001random, pedregosa2011scikit}) and assess performance across various spatial resolution scenarios. 
We note that we work on the raw PM2.5 sensor data collected acknowledging this as a potential source of limitation. Our modeling objective is the relative comparison of different urban areas in terms of PM2.5 levels. In S.I. (Section ~\ref{sec:supporting}) we discuss how the PM2.5 we employ is in alignment with other calibrated low cost sensor data and, moreover, when calibration techniques were applied for during testing, significant signal deviations have been rare.

\subsubsection{Problem Formulation}

For each city, we define a set of hexagonal areas, denoted as \(A\). Additionally, we consider monthly temporal snapshots with the goal of predicting the mean PM2.5 value observed at a given area in a given month \(t\). This prediction is based on a feature vector \(\mathbf{x}\), which comprises signals observed during the current month as well as historical data from the previous month \(t-1\). This setup represents a rolling prediction task, where the train-and-predict process is repeated monthly over a twelve-month period for each city.

In our experiments, we utilize three spatial resolution levels of the H3 index\footnote{\url{https://h3geo.org}}, a widely adopted discrete global grid system that segments geographies into hexagonal grids. The chosen resolution levels are 7, 8, and 9, corresponding to areas of \(5.16\), \(0.74\), and \(0.10\)  square kilometers, respectively. These levels enable us to evaluate model performance across a wide range of spatial aggregation levels, down to high-resolution areas of approximately 100 by 100 meters.

\subsubsection{Data and Features}

To construct the feature vector \(\mathbf{x}\), we incorporate various data sources aimed at capturing urban contextual characteristics, human activity, and the urban landscape. The underlying hypothesis is that factors such as traffic conditions, human mobility, street network features, and the characteristics of the built environment can explain the spatio-temporal variations in PM2.5 levels in urban areas. These data sources are grouped into four primary domains:

\begin{itemize}
    \item \textbf{Population Mobility:}
    Human mobility data serves as a proxy for activity levels within an area, based on the assumption that increased commercial and residential activity may correlate with higher PM2.5 levels. This data, derived from GPS pings sourced from mobile devices carried by humans, has been extensively used in human mobility research over the recent years~\cite{pepe2020covid, li2022location, wang2019extracting, aleta2020modelling, nande2021effect} . We define four signals in this category: \texttt{users\_counts\_{t-1}}, \texttt{ping\_counts\_{t-1}}, \texttt{user\_counts\_{t}}, and \texttt{ping\_counts\_{t}}, which represent aggregate, non-identifying, mobile user and GPS ping counts, in a hexagonal geographic area, during months  \(t-1\) and \(t\) respectively. In section~\ref{sec:datasets} we describe how the data has been processed in more detail. 

    \item \textbf{Built Environment:}  
    Features in this category capture various aspects of the built environment, including buildings, green spaces, parks, and street networks. The hypothesis is that elements such as extensive road networks or building infrastructure may contribute to higher pollution levels, while green spaces may help mitigate them. Using OpenStreetMap (OSM) data and the OSMNX~\cite{boeing2017osmnx} package, we extract the following features:  
    \begin{itemize}
        \item \texttt{green\_area\_points}: Number of green space POIs (e.g., parks).  
        \item \texttt{built\_area}: Total area ($km^{2}$) covered by building POIs.  
        \item \texttt{num\_buildings}: Number of building POIs in the area.  
        \item \texttt{street\_net\_density}: Number of street intersections.  
        \item \texttt{street\_net\_capacity}: Total area ($km^{2}$) covered by the street network.  
    \end{itemize}
    Larger road sections, for example, may indicate higher traffic volumes, which are often associated with increased PM2.5 levels.
    
    \item \textbf{Traffic:}  
    Vehicular GPS speed information (measured in km/h), has been  widely used as input to traffic estimation models in urban environments~\cite{herrera2010evaluation, meng2017city, yuan2021alternative}. To mode traffic congestion, we introduce the \texttt{speed\_avg\_t} feature, defined as the mean speed of all taxis operating within an area during the prediction month \(t\). Traffic conditions have been hypothesized as a contributor to PM2.5 levels in past studies~\cite{wen2024dynamic}, and this feature provides a time-varying indicator of such dynamics. 
    
    \item \textbf{Air Quality:}  
    We include \texttt{pm\_history\_{t-1}}, the mean PM2.5 level of the area during the previous month \(t-1\), to account for temporal stability and leverage historical trends of PM2.5 levels in the prediction task.

\end{itemize}

Among these categories, the \textit{Mobility}, \textit{Traffic}, and \textit{Air Quality} features are dynamic, with values varying monthly. In contrast, \textit{Built Environment} features are static, varying across areas but remaining constant over time.

\subsubsection{Experimental Setup}

\paragraph{\textbf{Evaluation Framework}}
For each prediction instance within a given city and month, we perform a leave-one-out cross-validation (LOOCV) prediction task. Specifically, we hold out the target prediction hexagonal area \( a \) and train supervised learning models using the training dataset \( X_{A'} \), where \( A' = A \setminus \{a\} \), which consists of data from all other areas in the city. $X_{A'}$ is made of a stack of all hexagonal area vectors $\mathbf{x}$ and corresponding target variables $\mathbf{y_{t}}$. LOOCV is then repeated for all areas within the city and mean for each performance metric is calculated. The cross-validation method employed is computationally expensive, as it requires training and testing the models \( n \) times, where \( n \) is the number of areas. However, it provides the advantage of yielding an unbiased estimate of the generalization error in a cross-validation setting~\cite{friedman2009elements}. 

\paragraph{\textbf{Models and Metrics}} 
In the prediction framework discussed, for each iteration in a given city and month $t$, a vector \( \mathbf{\hat{y}_t} \) is generated, representing the mean PM2.5 predictions for all hexagonal areas in \( A \). For evaluation purposes, these predictions are compared to the ground truth vector \( \mathbf{y}_t \). Using these vectors, we compute several evaluation metrics to assess model performance as a regression task, including Root Mean Squared Error (RMSE) and Spearman's rank correlation coefficient scores to assess model performance assuming a ranking task. Next, we report the results of a Random Forest Regressor trained  with \( 80 \) estimators and a maximum tree depth of 10 (we have experimented with a ridge and a decision tree regression model, but the Random Forests classifier achieved significantly better performance). 

\begin{figure*}[htbp]
    \centering
    \begin{subfigure}[b]{0.33\textwidth}
        \centering
        \includegraphics[width=\textwidth]{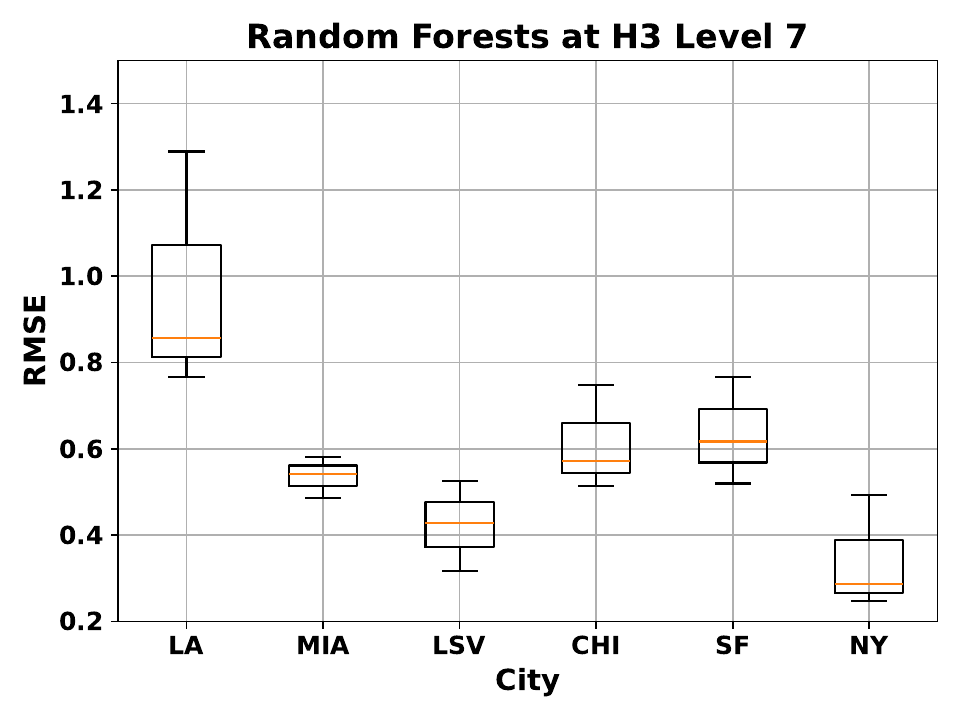}
        \caption{H3 Level 7}
        \label{fig:h3_p7}
    \end{subfigure}%
    \begin{subfigure}[b]{0.33\textwidth}
        \centering
        \includegraphics[width=\textwidth]{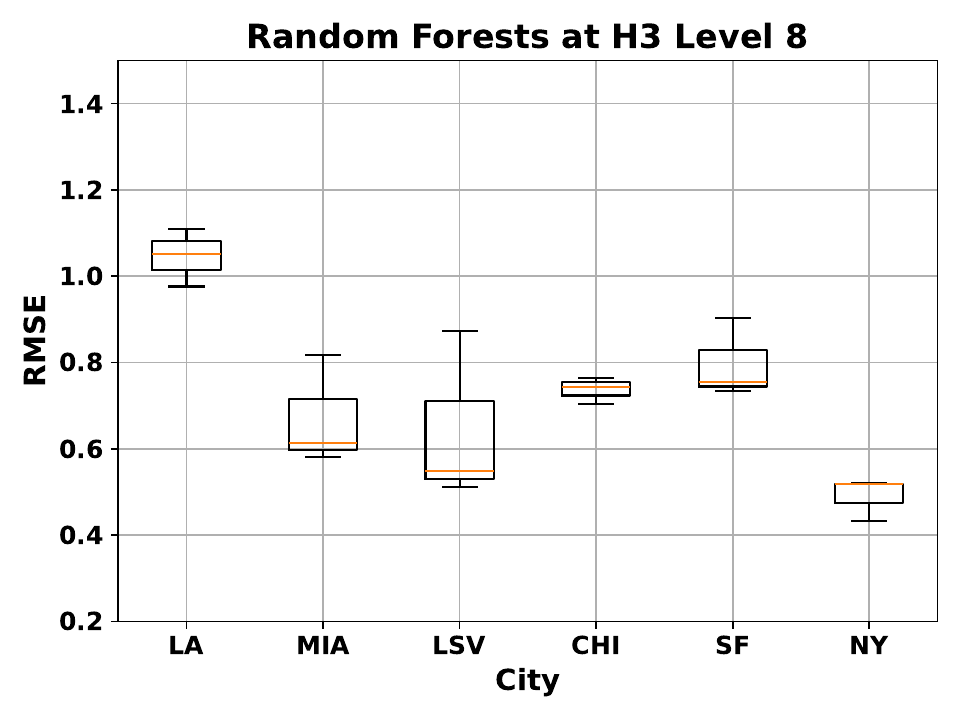}
        \caption{H3 Level 8}
        \label{fig:h3_p8}
    \end{subfigure}%
    \begin{subfigure}[b]{0.33\textwidth}
        \centering
        \includegraphics[width=\textwidth]{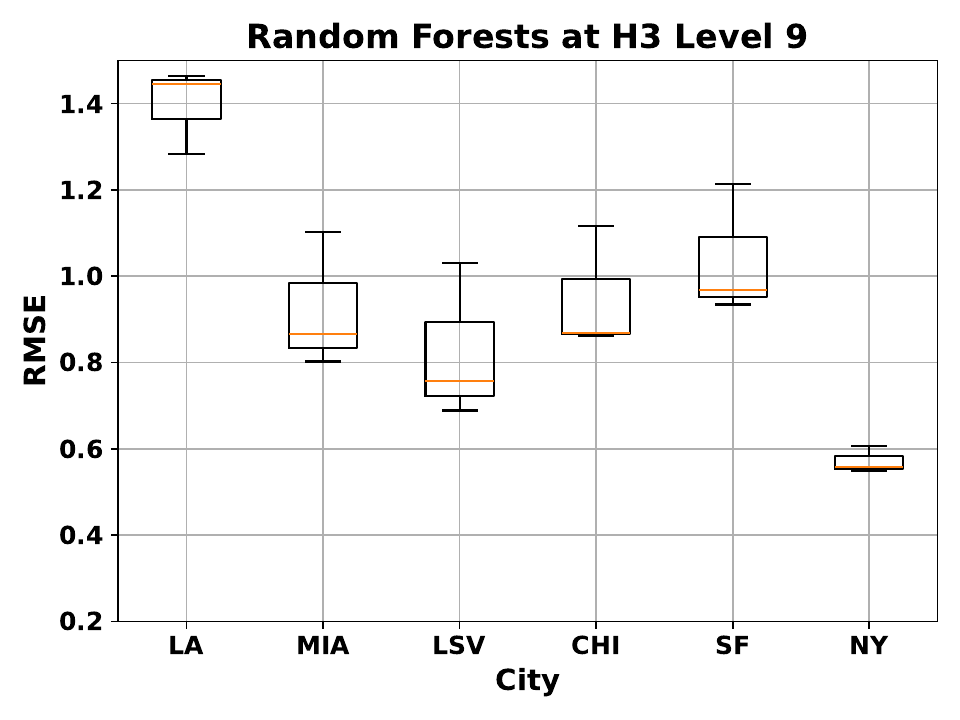}
        \caption{H3 Level 9}
        \label{fig:h3_p9}
    \end{subfigure}
    \caption{RMSE for Random Forest Predictions Across Different H3 Levels.}
    \label{fig:rmse_rf_boxplots}
\end{figure*}

\begin{figure*}[htbp]
    \centering
    \begin{subfigure}[b]{0.33\textwidth}
        \centering
        \includegraphics[width=\textwidth]{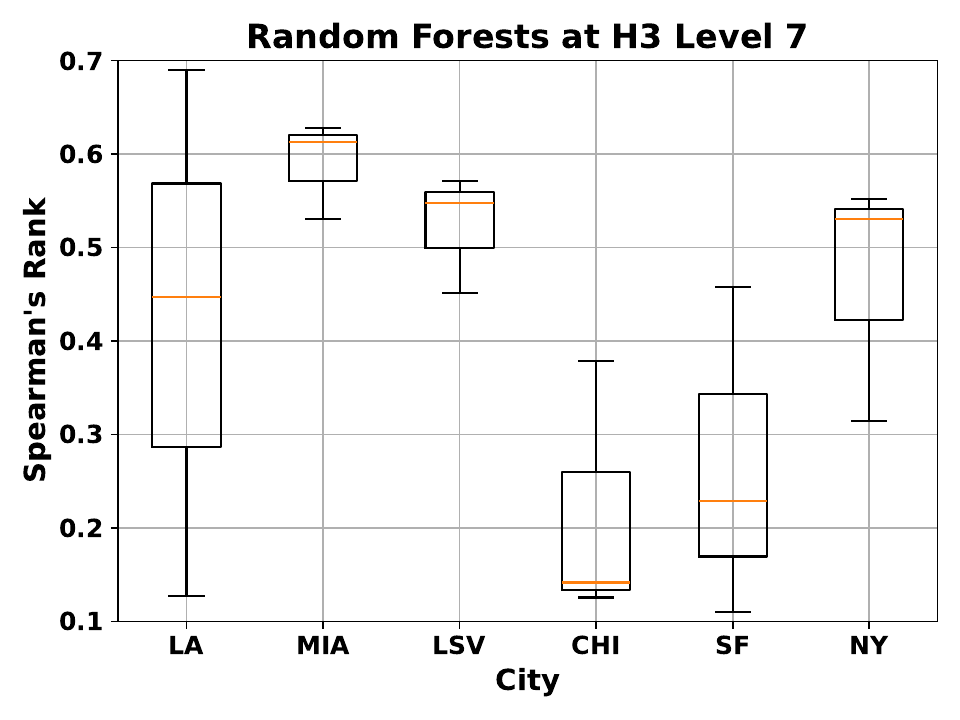}
        \caption{H3 Level 7}
        \label{fig:spearmanr_h3_p7}
    \end{subfigure}%
    \begin{subfigure}[b]{0.33\textwidth}
        \centering
        \includegraphics[width=\textwidth]{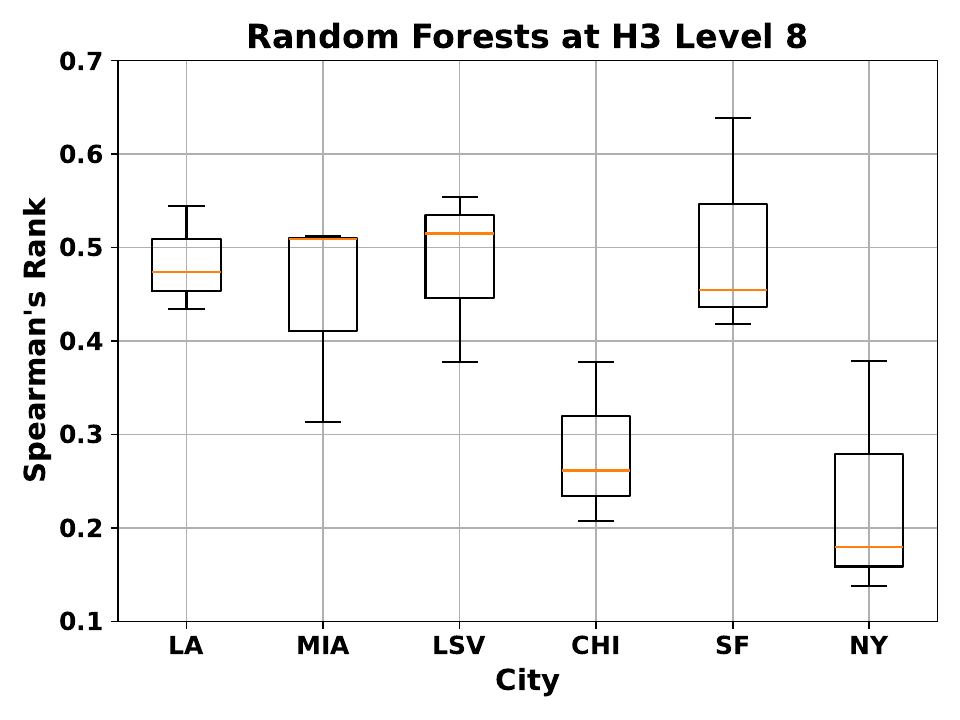}
        \caption{H3 Level 8}
        \label{fig:spearmanr_h3_p8}
    \end{subfigure}%
    \begin{subfigure}[b]{0.33\textwidth}
        \centering
        \includegraphics[width=\textwidth]{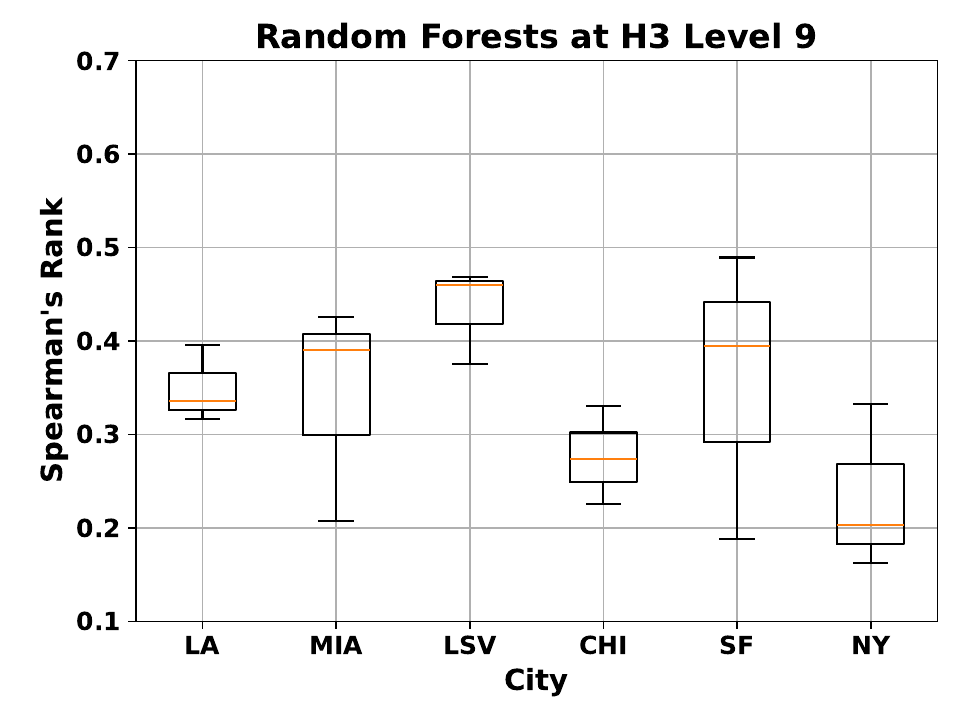}
        \caption{H3 Level 9}
        \label{fig:spearmanr_h3_p9}
    \end{subfigure}
    \caption{Spearman Rank Correlation for Random Forest Predictions Across Different H3 Levels.}
    \label{fig:spearmanr_rf_boxplots}
\end{figure*}

\subsubsection{Prediction Results}
\paragraph{\textbf{City level Performance}} The boxplots in Fig.~\ref{fig:rmse_rf_boxplots} (top row) illustrate the RMSE of Random Forest predictions across six cities (Los Angeles (LA), Miami (MIA), Las Vegas (LSV), Chicago (CHI), San Francisco (SF), and New York (NY)) for three H3 spatial resolution levels: Level 7 (coarsest), Level 8, and Level 9 (most granular). RMSE values \textit{increase} as the spatial resolution becomes finer, moving from H3 Level 7 to Level 9. This highlights the challenges of predicting PM2.5 at higher granularity, where local variations are harder to model accurately. Notably, LA and CHI show the most significant increases in RMSE at finer resolutions, particularly at H3 Level 9. In New York and Las Vegas the model achieves consistently relatively lower RMSE values across all H3 levels, potentially due to higher quality PM2.5 data collected in those regions where data density is considerably higher to that of other cities in the dataset (see S.I. Table~\ref{tab:city_pm25}). In this setting, more frequent sampling not only means a better historic PM2.5 predictor, but also more stable prediction outcomes as sample size increases is tied to better mean PM2.5 estimates.  

In Fig.~\ref{fig:spearmanr_rf_boxplots},  we show the ranking accuracy of the Random Forest model using Spearman's Rank Correlation. This metric assesses the model's ability to predict the relative ordering of PM2.5 levels across areas. In comparison to RMSE the most notable difference is the drop in performance when considering the city of New York and the relatively high performance in Los Angeles, Miami and Las Vegas especially. A potential explanation when contrasting these observations with the RMSE results noted above, is that the ranking task becomes less complex in cities where spatial coverage is limited, resulting to fewer areas candidate for ranking. 
In summary, evaluating prediction performance across different spatial resolutions and urban environments underscores the inherent trade-offs between model accuracy and the granularity of spatial representation for pollution levels. The observed variation in prediction accuracy across different spatial scales differs significantly from one city to another, reflecting diverse urban characteristics. Furthermore, PM2.5 concentrations and their predictability varies — not only across different cities but also within individual cities over time —, as indicated by the boxplot ranges. This variability highlights the considerable challenges associated with accurately modeling the spatiotemporal dynamics of pollutants. Addressing these challenges necessitates precise measurement methods combined with integrating diverse data sources into predictive modeling frameworks.

\begin{figure*}[htbp]
    \centering
    \begin{subfigure}[t]{1.0\textwidth}
        \centering
        \includegraphics[width=\textwidth]{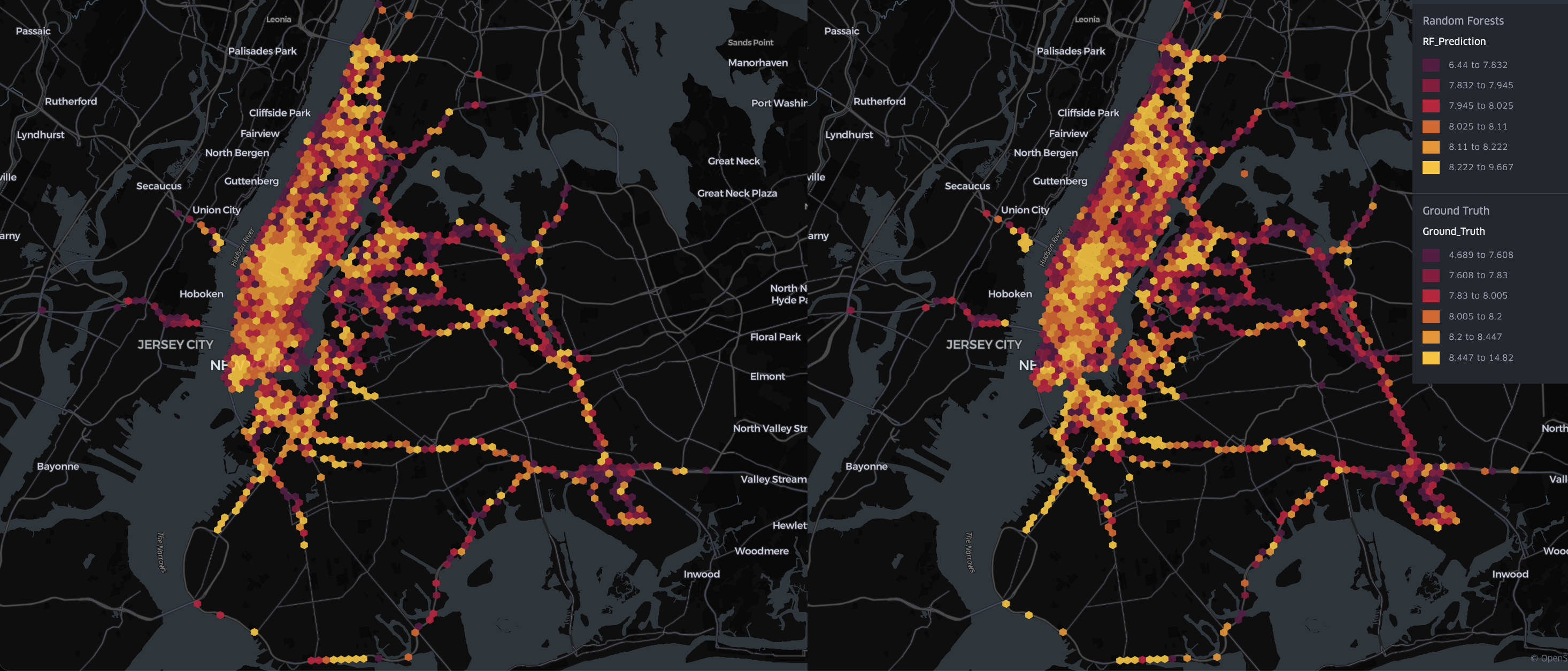}
        \label{fig:ny_kepler_comparison}
    \end{subfigure}
    \hfill
    \begin{subfigure}[t]{1.0\textwidth}
        \centering
        \includegraphics[width=\textwidth]{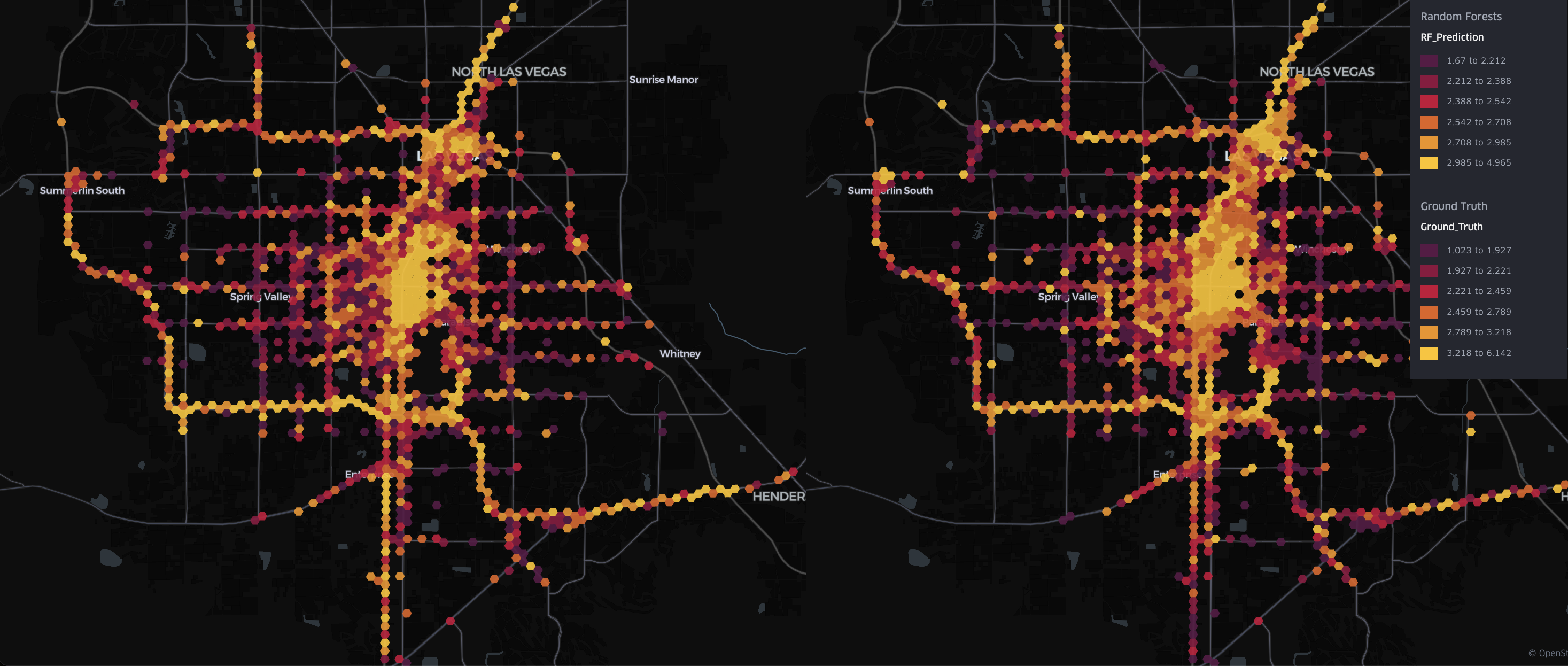}
        \label{fig:lv_kepler_comparison}
    \end{subfigure}
    \caption{Geographic area comparison of Random Forest mean PM2.5 predictions (left) varsus ground truth data (right) for New York City and Las Vegas.}
    \label{fig:kepler_comparison}
\end{figure*}
\paragraph{\textbf{Spatial Comparison of Predictions with Ground Truth Data}}
Next, we inspect prediction performance visually, aiming to provider a richer as well as a more intuitive perspective on PM2.5 predictability within cities. 
Figure~\ref{fig:kepler_comparison} provides a visual comparison of the spatial distribution of PM2.5 predictions generated by the Random Forest model and the ground truth PM2.5 data  for New York City and Las Vegas, respectively. The visualization allows for an assessment of the model’s ability to replicate spatial patterns of PM2.5 concentrations.
In New York, predictions exhibit a strong spatial agreement with the ground truth data, particularly in high-density urban areas such as Manhattan and parts of Brooklyn. Performance in central Manhattan and major traffic corridors, highlights the model’s ability to leverage urban activity and mobility-related features in predicting PM2.5 levels. Discrepancies are observed in peripheral areas, such as parts of Staten Island and the outskirts of Queens, where PM2.5 levels tend to be lower. The model occasionally fails to fully capture the subtle variations in air quality in these regions. In Las Vegas the models also shows good alignment with the ground truth data particularly in central urban areas, along major roads and highways. 
All in all, the spatial distribution of the Random Forest predictions demonstrates a strong alignment with ground truth data in high-density and high-pollution regions. However, discrepancies in peripheral and suburban areas, highlight the model's limitations in capturing subtle variations in less dense regions.  These results suggest that while the model performs well in urban cores and areas with strong feature representation, additional refinements may be necessary to improve predictions in less data-rich environments such as the peripheral areas of the city. We discuss modeling implications around data equity in Section~\ref{sec:discussion}.
\begin{figure*}[htbp]
    \centering
    \includegraphics[width=\textwidth]{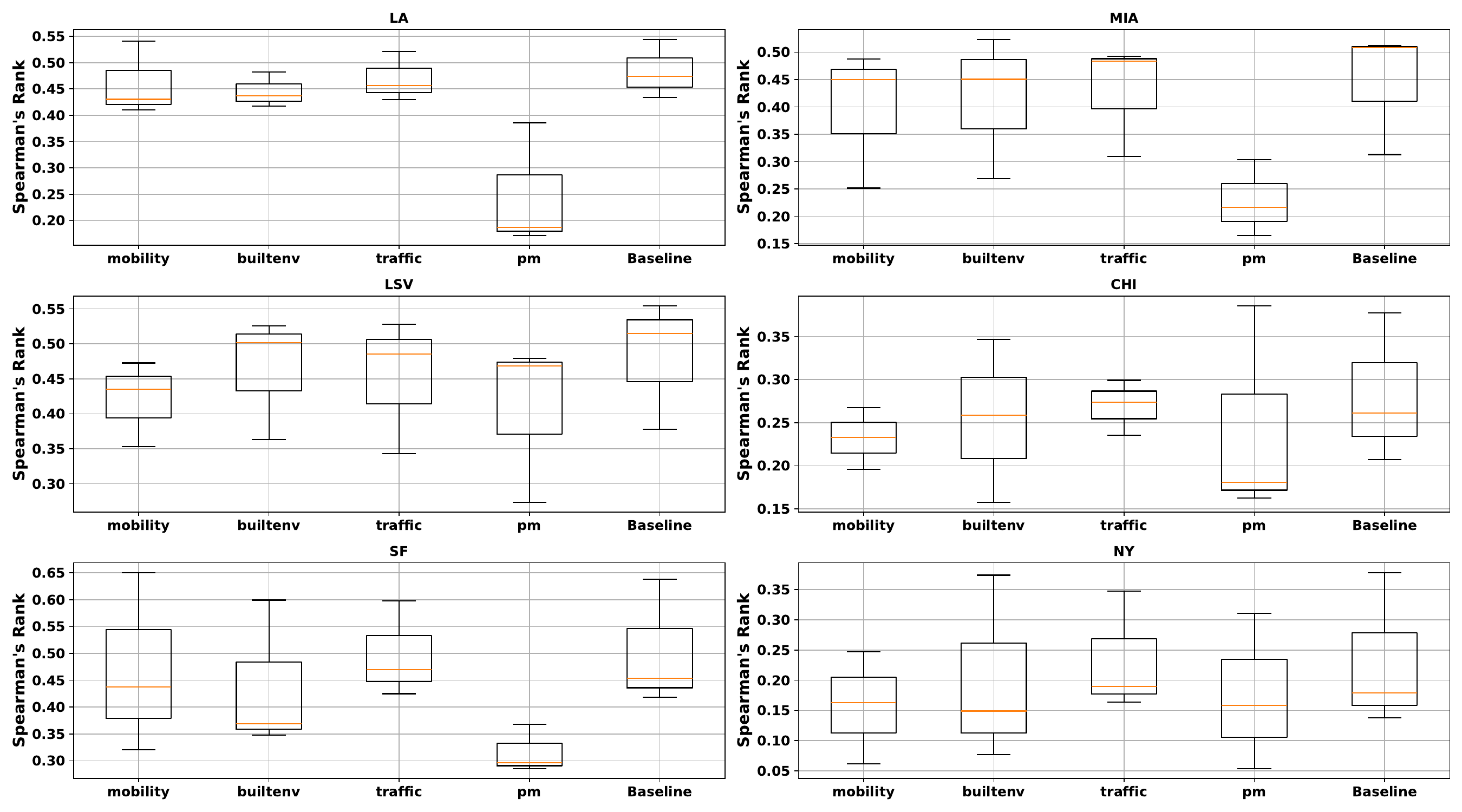}
    \caption{Feature drop analysis for Spearman Rank Correlation of Random Forest predictions at H3 Level 8.}
    \label{fig:feature_drop_spearmanr_h3_p8}
\end{figure*}
\paragraph{\textbf{Feature Importance Analysis}} Figure~\ref{fig:feature_drop_spearmanr_h3_p8} presents a feature drop analysis for the Spearman Rank Correlation of Random Forest predictions at H3 Level 8 across the six cities. The analysis evaluates the impact of different feature categories on the model’s performance by removing each category and observing the resulting Spearman Rank Correlation.
For all cities, the baseline model that includes all feature categories achieves the highest Spearman Rank Correlation, demonstrating the value of combining diverse feature sets and the multi-factorial origin of PM2.5 levels. Moreover, the drop in performance when removing any feature category points in the same direction. More specifically, removing Air Quality (PM) historic features leads to the largest decline in Spearman Rank Correlation for most cities, most notably in \textit{Los Angeles (LA)}, \textit{San Francisco (SF)}, and \textit{Miami (MIA)}. This underscores the importance of regularly performing air quality data sampling locally to accurately capture the general tendency of an area to feature certain levels of PM2.5.  Mobility features appear also to be a key source of information towards predicting urban air quality levels. Its impact is larger in the cities with larger population density such as New York and Chicago. Highlighting how higher population density is indicative of an environment that is prone to lower air-quality standards as human activity nearby intensifies. The built environment feature set, which integrates information on green area coverage, as well as street network and area construction characteristics appears more essential to higher levels of predictability in Chicago, New York and San Francisco. In those cities, its removal not only shows the greatest amount of variability over time, but also the highest drop in performance. It is worth noting that despite being a set of features comprised of temporally static indicators, its performance demonstrates how urban environment descriptors are still valuable proxies to the capacity of an area to be a host of pollutants at a certain level. Among the six cities, Chicago, New York and San Francisco have a considerably higher density of medium and high rise buildings. The morphology of urban areas significantly influences local air quality, particularly within densely built environments where street canyons are common. These canyons, formed by continuous rows of buildings, restrict natural airflow patterns and can severely hinder the dispersion of atmospheric pollutants \cite{Hang2012}. The effect is often exacerbated in districts characterized by high-rise buildings, which create deep canyons with high aspect ratios (height/width). Such geometries have been shown to impede ventilation efficiency, reducing the capacity for pollutant removal and potentially trapping emissions near ground level \cite{He2017}.

Traffic as a signal for prediction appear less important in discriminating across areas of a city in terms of their PM2.5 levels, though it still adds value to the prediction task and its variability in performance suggests that in certain occasions it can be a critical source of information to rely on. We demonstrate how it can be essential in modeling PM2.5 variations across cities and seasonally, when we present a macroscopic viewpoint on PM2.5 modeling in the following section. Overall, these findings suggest that the predictive utility of different feature categories is highly context-dependent, highlighting the need to tailor feature engineering to the unique characteristics of each city. The latter insight becomes especially relevant when taking into account that data quality across all set of signals can vary from one urban environment to another.


\begin{figure*}
  \centering
  \begin{subfigure}[b]{0.32\columnwidth}
    \centering
\includegraphics[width=\textwidth]{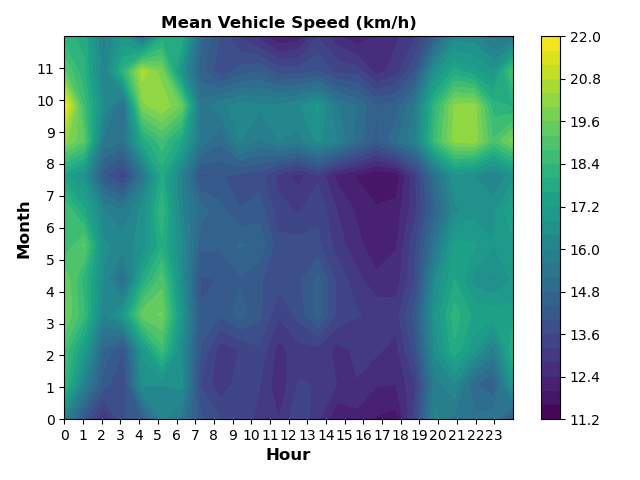}
    \caption{}
    \label{fig:mean_speed_CHI}
  \end{subfigure}
  \hfill
  \begin{subfigure}[b]{0.32\columnwidth}
    \centering
\includegraphics[width=\textwidth]{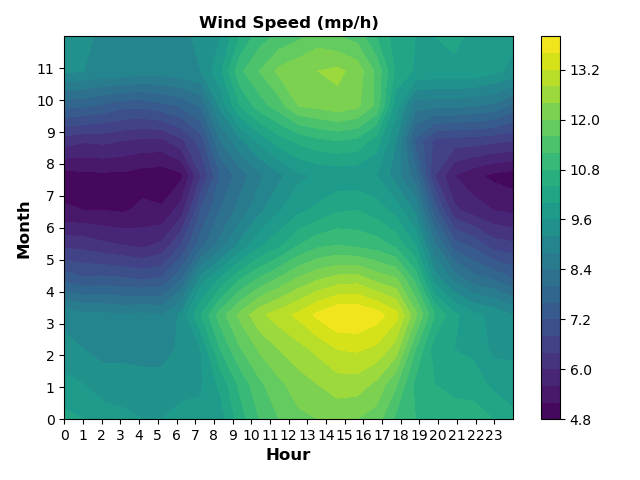}
    \caption{}
    \label{fig:wind_speeds_CHI}
  \end{subfigure}
  \hfill
  \begin{subfigure}[b]{0.32\columnwidth}
    \centering
\includegraphics[width=\textwidth]{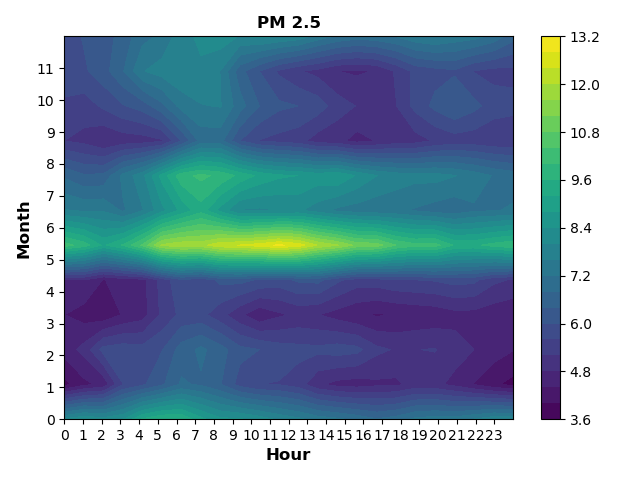}
    \caption{}
    \label{fig:pm_CHI}
  \end{subfigure}

  \begin{subfigure}[b]{0.32\columnwidth}
    \centering
\includegraphics[width=\textwidth]{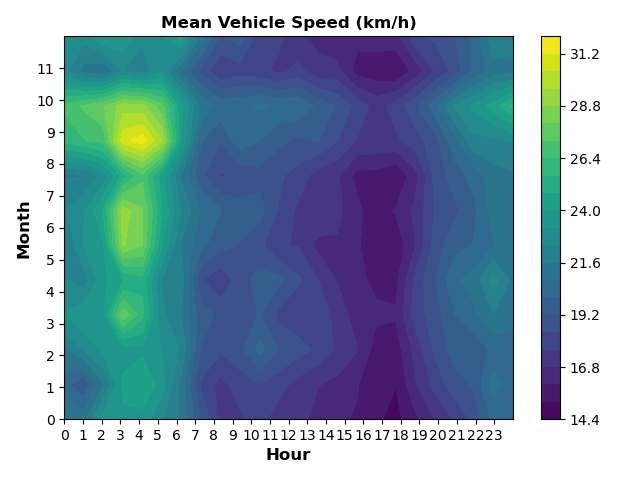}
    \caption{}
  \end{subfigure}
  \hfill
  \begin{subfigure}[b]{0.32\columnwidth}
    \centering
    \includegraphics[width=\textwidth]
    {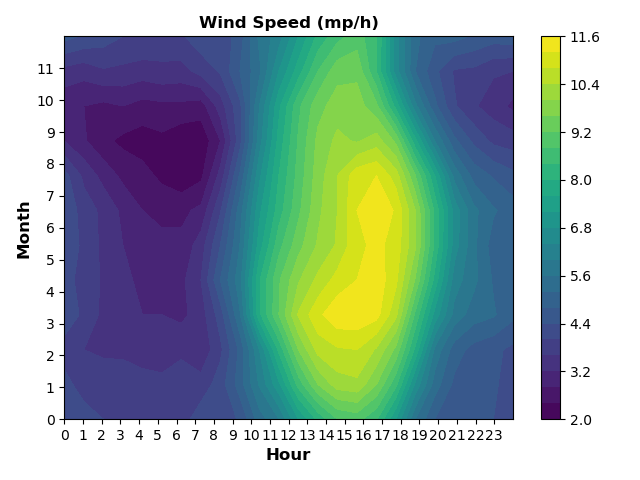}
    \caption{}
    \label{fig:wind_speeds_LA}
  \end{subfigure}
  \hfill
  \begin{subfigure}[b]{0.32\columnwidth}
    \centering
    \includegraphics[width=\textwidth]{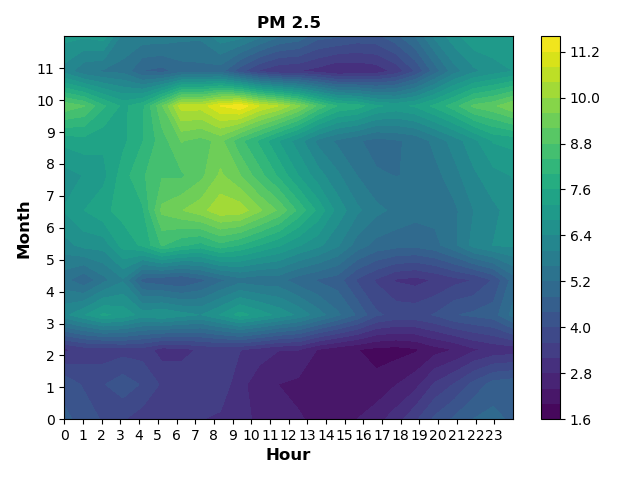}
    \caption{}
    \label{fig:pm_LA}
  \end{subfigure}
  
  \caption{Contour plots representing intensity levels of mean vehicle speed (km/h), mean wind speed (km/h) and mean PM2.5 levels in the Chicago (top) and Los Angeles areas (bottom). Signal variations are shown for different months (1-12) and daily hours (0-23).}
  \label{fig:overall}
\end{figure*}

\begin{figure*}[htbp]
  \centering
  \begin{subfigure}[b]{0.48\textwidth}
    \centering
    \includegraphics[width=\textwidth]{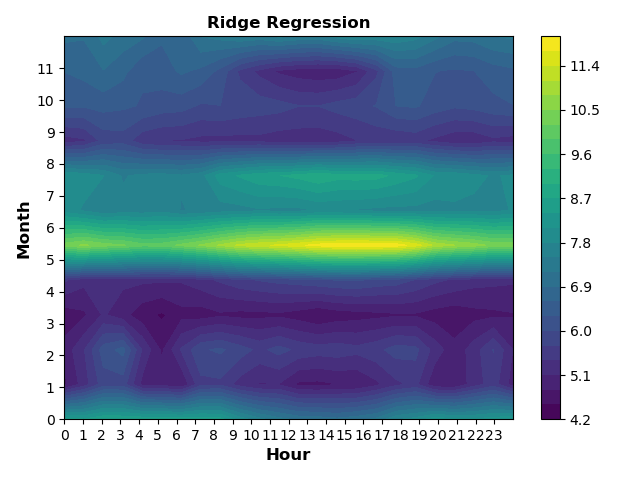}
    \caption{Chicago}
    \label{fig:modelchi}
  \end{subfigure}
  \hfill
  \begin{subfigure}[b]{0.48\textwidth}
    \centering
    \includegraphics[width=\textwidth]{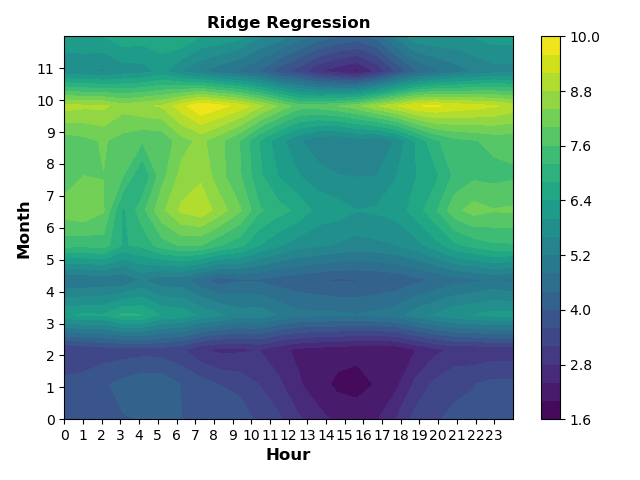}

    \caption{Los Angeles}
    \label{fig:modella}
  \end{subfigure}
  \caption{Congestion-Wind Regression Model Output in the Chicago and Los Angeles Areas when considering a monthly training dataset.}
  \label{fig:modeloutputs}
\end{figure*}

\begin{figure*}[htbp]
  \centering
  \begin{subfigure}[b]{0.48\textwidth}
    \centering
    \includegraphics[width=\textwidth]{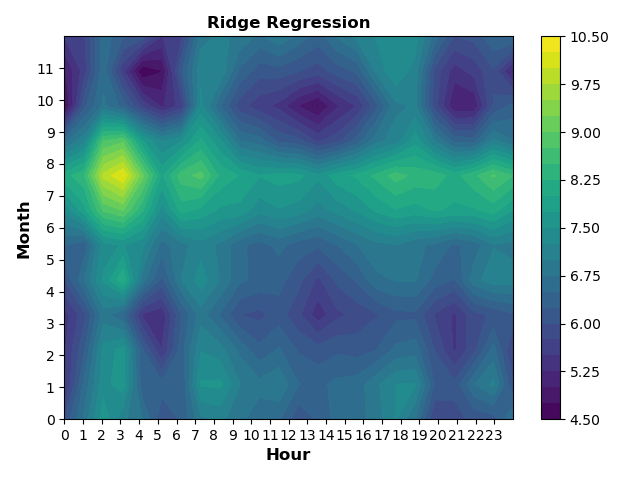}
    \caption{Chicago}
    \label{fig:modelchi}
  \end{subfigure}
  \hfill
  \begin{subfigure}[b]{0.48\textwidth}
    \centering
    \includegraphics[width=\textwidth]{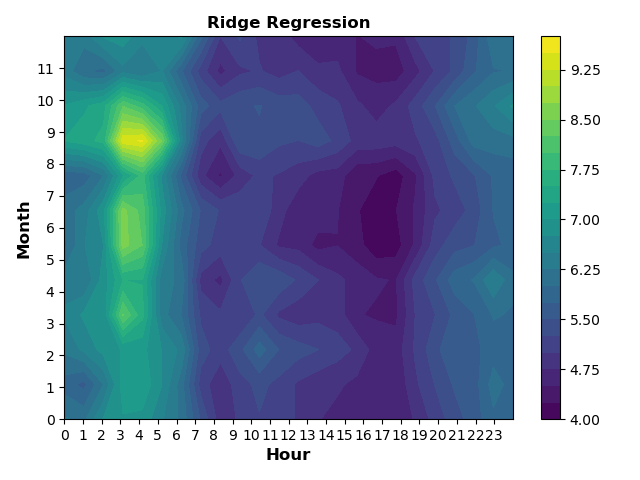}

    \caption{Los Angeles}
    \label{fig:modella}
  \end{subfigure}
  \caption{Congestion-Wind Regression Model Output in the Chicago and Los Angeles Areas when considering an annual training dataset.}
  \label{fig:modeloutputs_annual}
\end{figure*}

\subsection{Macroscopic Seasonal Modeling of PM2.5}
\label{sec:macro}
\begin{figure}[ht]
    \centering
    \includegraphics[width=\columnwidth]{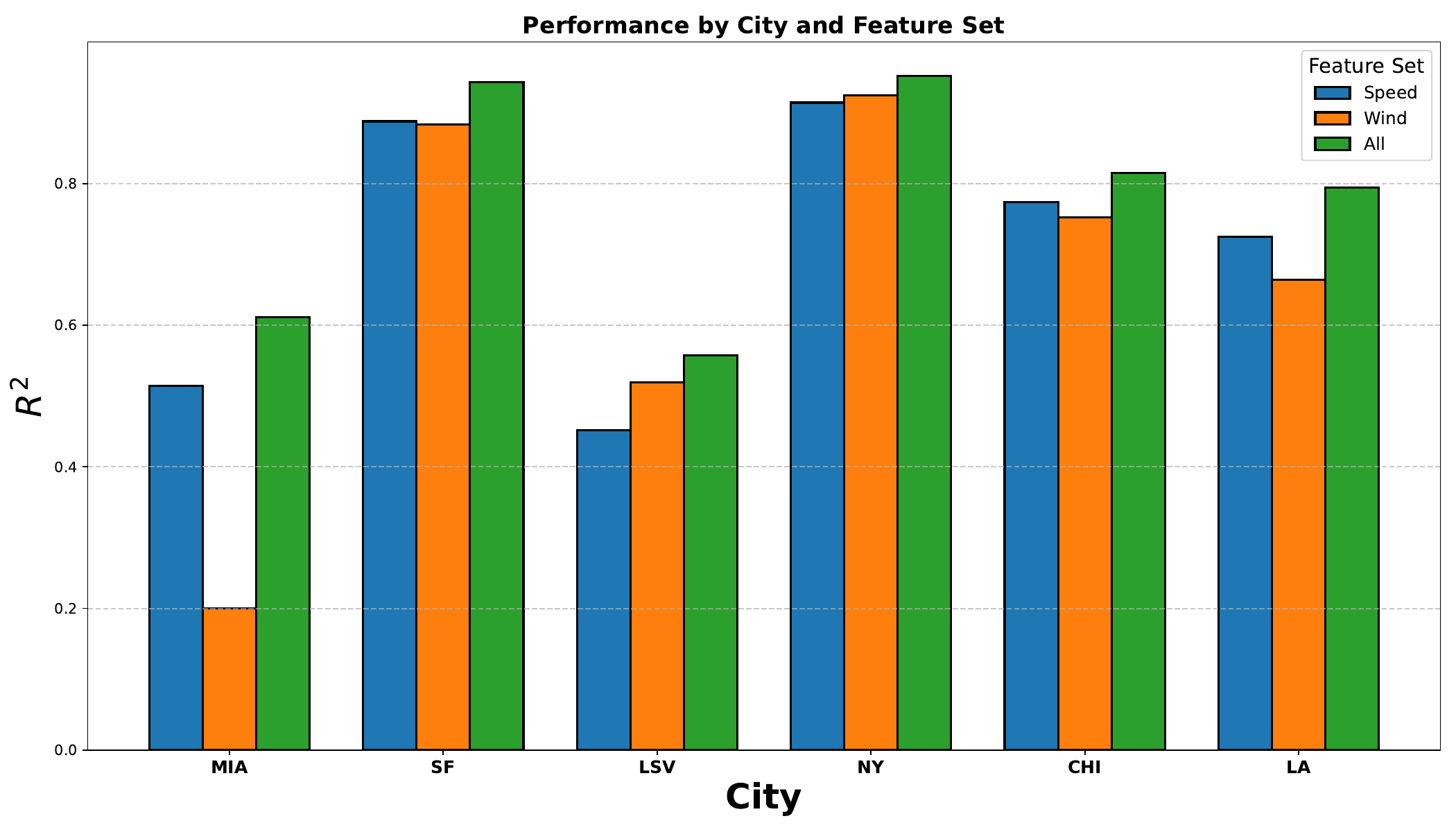}
    \caption{Coefficient of determination score $R^2$ by City and Feature Set for the monthly-trained regression model.}
    \label{fig:macro_feature_performance_R2_True}
\end{figure}

Our next goal is to model macroscopic PM2.5 patterns, focusing on seasonal variations in urban air quality and comparing observed patterns across cities. In this case, we adopt an epistemological approach of model minimalism and interpretability, over accuracy and hyperlocal predictions seen in the previous section. In this setting, we are also able to investigate diurnal variations in PM2.5 levels, a dimension which has been challenging to examine in small city areas, due to data sparsity.

Wind patterns have been previously hypothesized to play a key role in the dispersion and concentration of pollutants~\cite{ferm2015concentrations}, 
though the specific dynamics of the relationship between PM2.5 and wind vectors remain a topic of ongoing research. On the one hand, strong winds can dilute PM2.5 concentrations introducing effectively a ventilation effect in a city. However, the impact of wind patterns on air quality can also depend on the presence of proximate industry sites that can become pollution sources for a city, a case where wind direction in particular can play a significant role~\cite{zhang2021using}, but also natural phenomena e.g. forest fires or dust winds, where  particles can move across large geographic distances~\cite{o2019contribution}. With these into consideration, we utilize publicly available data on hourly wind speed levels available for the regions of the cities we study here~\cite{climate_normals, arguez2010noaa}. 
While we operate under the assumption that winds follow seasonal patterns without dramatic year-to-year shifts based on a 30-year historical average (1990-2020), it is important to acknowledge the limitations of this approach. Inter-annual variability, driven by large-scale climate oscillations such as the North Atlantic Oscillation \cite{hurrell_decadal_1995} and Pacific Decadal Oscillation \cite{trenberth_decadal_1994}, can indeed lead to significant deviations from these averages in any given year. Despite these potential fluctuations, utilizing a 30-year average, consistent with guidelines from the World Meteorological Organization \cite{wmo_guidelines_2017}, provides a standard climatological baseline, often referred to as the wind normal in the United States, for understanding general wind behavior and resource assessment.

Figures~\ref{fig:wind_speeds_CHI}~and~\ref{fig:wind_speeds_LA} we visualize the mean wind speed aggregated over a twelve month time window in the Chicago and Los Angeles areas. We provide similar hourly seasonal contour plots for two more variables that are the subject of the present study, namely mean car speed as collected by the taxi sensors in the dataset, and PM2.5 levels, as depicted in see Fig. \ref{fig:overall}, (results for all cities are available in S.I.). We clarify that wind data was not used in the previous section as it becomes available only at the city level and not at the neighborhood level. 

By inspecting the figures the reader can make a few key observations. Firstly, seasonal and diurnal PM2.5 patterns can vary from city to city. In other words, each city presents its own idiosyncratic form of air pollution levels. Secondly, PM2.5 concentrations at a given month and hour are typically at lows when wind speeds are at high levels. Thirdly, at times in the year where congestion levels are low (mean speed is high), PM2.5 levels are typically lower. Wind speed and congestion alone do not appear to match the temporal evolution patterns of PM2.5 levels. We build a regression model  that exploits the combination of the two information signals to predict the temporal evolution patterns of PM2.5 for all cities in the dataset. In this instance, we have elected a
ridge regression model~\cite{breiman2001random, pedregosa2011scikit} with a regularization parameter \( \gamma \) set to \( 0.1 \). Informal experiments, with non linear models like decision trees and random forests ensemble have shown inferior performance in this task, failing to model the smooth temporal transitions of diurnal and seasonal variations observed in the data.

In Figure~\ref{fig:modeloutputs}~and~Figure~\ref{fig:modeloutputs_annual} we present the results of the model, trained using two training data set ups, monthly and annual respectively. In both scenarios we remove the data point under prediction, i.e. that corresponds to the test month $m_t$ and hour $h_t$ during training, applying the leave-one-out cross validation technique described in the previous section.  
In the case of the monthly output we use as training data only data from the specific month we are aiming to predict PM2.5 levels at a given month $m$ and hour $h$, whereas the annual model uses data from all months available in the year. The results for the model trained on monthly data suggest that using only car and wind speed information, the characteristic patterns of PM2.5 levels about a city can be recovered to a large degree. This is a promising finding as traffic congestion information as well as wind speed information is more widely available than PM2.5 sensing data. The PM2.5 signal is still very useful nonetheless, as it has been employed as a label source in the supervised learning regression model dictating how wind and speed information interact to obtain PM2.5 estimates. 

Furthermore, despite using a smaller volume of training data, the monthly-specific model outperforms the model trained on the annual dataset. This may be an indication that the
interplay between the primary proxy of human activity (e.g. traffic) and meteorological indicators (e.g. wind) can be
season-specific and models that take seasonality into account are better fit to capture urban air quality patterns. Finally, Fig.~\ref{fig:macro_feature_performance_R2_True} shows  model performance in terms of coefficient of determination $R^{2}$ for different feature sets. Best performance is consistently observed across all cities in the dataset, when both wind and congestion patterns are taken into account. The capacity of models to capture PM2.5 patterns presents strong variations across cities.

\section{Discussion}
\label{sec:discussion}
\paragraph{Predictability of PM2.5 inside the city.}
The modeling outcomes presented in section~\ref{sec:modeling} underscore both the promise and complexity of fine-grained urban air quality prediction using heterogeneous, multi-source data. While the Random Forest model achieved strong alignment with ground truth PM2.5 levels—particularly in dense urban cores and along major transportation corridors—performance deteriorated at finer spatial resolutions and in peripheral areas. This degradation reflects the challenge of capturing localized pollution dynamics in data-sparse or spatially heterogeneous environments. Crucially, the predictive power of the machine learning model was highly dependent on the availability of historical PM2.5 measurements and mobility indicators, reinforcing the importance of temporal continuity and human activity signals in estimating pollution levels. Feature drop analysis also highlighted that static features of the built environment, such as green space, building density, and road networks, can provide meaningful structural context despite their temporal invariance. However, the relative predictive utility of each feature domain varied significantly across cities, indicating that urban morphology, infrastructure complexity, population characteristics, and data coverage deeply influence model performance. These results suggest that generalizable models cannot rely on uniform assumptions, but instead require context-aware, locally calibrated frameworks capable of adapting to unique urban signatures. 

Importantly, our findings reveal an underlying issue of data equity in urban sensing. Cities and neighborhoods with higher sensor density and richer mobility or infrastructure data benefited from more accurate predictions, while under-monitored areas—often lower-income or less digitally connected—saw higher error rates. This asymmetry raises critical concerns for data-driven policymaking: environmental models risk reinforcing existing disparities if they disproportionately perform well only in data-rich regions. As such, there is a compelling need for intentional investment in equitable data infrastructure, ensuring that high-quality PM2.5 monitoring extends beyond urban cores to include peripheral and under-served communities where perhaps taxi mobility is more limited. From a policy and operational standpoint, our study emphasizes the indispensable role of regular and reliable PM2.5 sensing. While proxy features—like mobility and congestion—are useful, their predictive utility depends on being anchored in recent, location-specific ground truth data. Models can only perform well where there is sufficient historical signal to learn from; without a stable foundation of air quality observations, even the most sophisticated predictive frameworks face significant limitations. Ensuring widespread, temporally consistent, and publicly accessible air quality data is therefore not just a scientific imperative, but a prerequisite for scalable and equitable environmental governance.

Recent studies have demonstrated the potential of machine learning for predicting PM2.5 at fine spatial scales using multi-source data, while also underscoring the need for contextual adaptation to different urban morphologies~\cite{liu2024fine}. The integration of low-cost sensor networks into predictive frameworks has further improved PM2.5 exposure assessment in areas with limited regulatory monitoring~\cite{elshorbany2023lowcost}. However, research has highlighted significant disparities in air quality data infrastructure across cities and neighborhoods, with communities of color often underrepresented in regulatory monitoring networks~\cite{mujahid2024disparities}. These findings reinforce calls for community-based sensing efforts and equitable investment in air quality infrastructure to close persistent data gaps~\cite{georgetown2023equity}. Eventually, governmental efforts, citizen-led projects and corporate platforms have to come in synergy so as to offer optimal air quality monitoring outcomes in urban ecosystems.  

A key limitation of our approach in using machine learning to predict PM2.5 locally concerns feature engineering. Although we have proposed a set of features derived from mining a broad set of data sources, there are multiple alternative formulations that could be evaluated. We envision that, through making the available data to the wider research community, improved versions of features and models can emerge, further pushing the boundary of prediction performance we have observed here. Modern deep learning techniques can relax, to an extent, the constraints around feature engineering, yet these models can also be data demanding. Combining data collected from sensors with satellite datasets can be a promising step forward in this direction~\cite{arafat2020deep}. The effectiveness of features is intrinsically linked to the quality of the underlying data as well. For example, the quality of OpenStreetMap data relies on the number of crowdsourced contributions~\cite{sehra2017assessing}, which can vary substantially across different regions. Similarly, the quality of mobility data depends on the penetration levels of service providers in various areas. With regards to the PM2.5 data, the stability of mean estimates improves as the number of samples increases for a given spatio-temporal snapshot. 

\paragraph{Seasonal PM2.5 Modeling}
In Section~\ref{sec:macro}, we approached the problem of predicting PM2.5 from a different perspective; we shifted away from a framework featuring a highly granular spatial resolution component, and focused on temporal granularity and seasonal patterns of PM2.5 levels at the city level. In this setting, we showed how models trained on monthly data outperform those trained on full-year datasets in capturing PM2.5 trends, despite having access to less training data overall. This counterintuitive outcome can be attributed to multiple seasonally dynamic processes that influence aerosol concentrations and urban air quality more broadly. Firstly, vertical mixing strength, which determines how pollutants disperse in the atmosphere, is known to vary significantly by season. During colder months, the planetary boundary layer is typically more stable and shallow, trapping pollutants near the surface. In contrast, warmer months encourage deeper mixing and greater dispersion, leading to lower pollutant concentrations under similar emission loads \cite{seinfeld2016atmospheric, holzworth1967, petaja2016}. Secondly, emissions themselves are not static year-round. Winter is often marked by increased use of residential heating, wood burning, and other combustion sources, which can drive up PM2.5 levels. During the summer, biogenic emissions and sunlight-driven chemical reactions become more influential, contributing to the formation of secondary organic aerosols through oxidation processes \cite{hallquist2009, goldstein2007}.

Another important seasonal effect is the temperature-dependent volatility of aerosol components. For instance, ammonium nitrate, a common secondary inorganic aerosol, tends to evaporate more readily in warm conditions, resulting in lower PM2.5 mass even when precursor gases are abundant \cite{seinfeld2016atmospheric}. These chemical and physical processes are not adequately captured by models that aggregate data across all seasons. This helps explain why monthly models, despite being data-sparse, are more effective: they implicitly capture the underlying seasonal context—ranging from emission profiles to atmospheric chemistry and transport. Our findings suggest that season-aware models using proxy variables such as traffic congestion and wind speed can be highly effective in reconstructing urban air quality dynamics across diverse geographic contexts. Critically, there are strong variations in the seasonal manifestation of PM2.5 patterns across geographic regions. Each city presents its own unique climatic, urban morphological and human activity characteristics that give rise to a distinct footprint of diurnal and seasonal leveling of PM2.5.

\section{Acknowledgments} 
A.N. and Y.A. thank the Firefly Product, Engineering and Operations teams for their coordinated efforts to instrument the data collection efforts amongst a nation-wide taxi fleet of thousands of vehicles.  R.L. is supported by the EPSRC grants EP/V013068/1, EP/V03474X/1 and EP/Y028872/1. 

\section{Data Availability}
\label{sec:datasets}
The PM2.5 and traffic congestion datasets used in this work are becoming  available to the research community through Firefly's Air Quality Data Licensing program\footnote{\url{https://www.fireflyon.com/air-quality}}. The OpenStreetMap street network characteristics as well the POI data are available through the open source OSMNX project~\cite{boeing2017osmnx}. The wind data can be obtained by accessing the openly available archive of U.S. Climate Normals\footnote{\url{https://www.ncei.noaa.gov/data/normals-hourly/1991-2020/}}. The mobility data used in the present work is proprietary\footnote{\url{www.unacast.com}} and access to it was provided to data through highly secure computing infrastructure operated by Firefly.  The data has been de-identified to guarantee user privacy, aggregated at the area level, while also undergoing additional privacy enhancements, such as the removal of certain POI categories and the obfuscation of home areas. While we cannot make mobility data available directly, researchers can obtain similar datasets through licensing programs including Cuebiq's Data For Good\footnote{\url{https://cuebiq.com/social-impact/}} initiative which has become widely used by the research community~\cite{li2022location, wang2019extracting, aleta2020modelling, nande2021effect}.

\section{Supporting Information}
\label{sec:supporting}
\begin{figure}[]
  \centering
 \includegraphics[width=0.9\columnwidth]{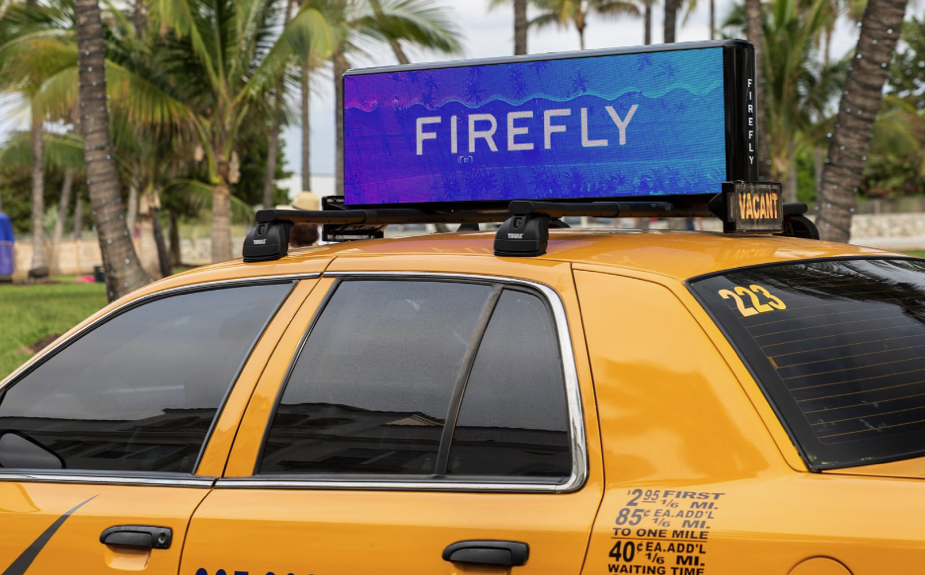}
  \caption{Firefly Taxi top in operation in Miami.}
  \label{fig:sunrise}
\end{figure}

Figure~\ref{fig:sunrise} shows Firefly's Sunrise taxi top. The display features an Android board for processing and is equipped with a set of sensors including a GPS sensor. The air quality sensor important to  this study has been laser-based particulate matter sensor (MFG: BYD, PN: 2006H) with dimensions of 44~$\times$~31.6~$\times$~10.5~mm. It operates over a wide temperature range, with a working temperature range of $-20\,^{\circ}\mathrm{C}$ to $70\,^{\circ}\mathrm{C}$ and a storage temperature range of $-40\,^{\circ}\mathrm{C}$ to $85\,^{\circ}\mathrm{C}$. The sensor is capable of detecting particulate matter within the size range of PM0.3 to PM10, with a sensitivity of $\pm 15\,\mu\mathrm{g/m}^3$ in the concentration range of 0 to $100\,\mu\mathrm{g/m}^3$. The use of laser-based detection ensures high accuracy and responsiveness for monitoring ambient particulate levels. These sensors can be sensitive to humidity effects and appropriate calibration methods have been developed to account for potential measurement biases as we explain in the following paragraphs.

\subsection{Firefly PM2.5 Sensing Comparison With Other Sources}
\paragraph{EPA}
\begin{table}[htbp]
  \centering
\begin{tabular}{lllll}
\toprule
City & Data points & Mean & SD \\ 
\midrule
New York & 42,272,355.00 & 7.58 & 11.82 \\ 
Chicago & 16,557,911.00 & 7.45 & 8.73 \\ 
Los Angeles & 15,761,095.00 & 6.38 & 8.71 \\ 
San Francisco & 13,418,809.00 & 6.22 & 8.07 \\ 
Miami & 12,067,138.00 & 4.73 & 6.20 \\ 
Las Vegas & 37,189,840.00 & 2.64 & 6.83 \\ 
\bottomrule
\end{tabular}
\caption{City level PM2.5 basic statistics, Firefly Taxis}
\label{tab:city_pm25}
\end{table}
\vspace{+20pt}
\begin{table}[htbp]
  \centering
\begin{tabular}{ll}
\toprule
City &  PM2.5\\
\midrule
Los Angeles-Long Beach-Anaheim, CA & 12.1 \\
Las Vegas-Henderson-Paradise, NV & 10.5 \\
Chicago-Naperville-Elgin, IL-IN-WI & 10 \\
San Francisco-Oakland-Hayward, CA & 9.9 \\
Miami-Fort Lauderdale-West Palm Beach, FL & 9.4 \\
New York-Newark-Jersey City, NY-NJ-PA & 8.7 \\
\bottomrule
\end{tabular}
\caption{Annual mean PM2.5 values reported by the US EPA in similar regions (measured in µg/m3).}
\label{tab:EPA_pm25}
\end{table}

We compare PM2.5 levels reported by taxis equipped with low cost sensors with levels reported by the Environment Protection Agency (EPA) in the U.S. across very broad geographic areas.  Key factors that come into play during these comparative analysis include spatio-temporal sampling rate considerations across the various monitoring platforms as well as regional spatial boundary definitions. 
In Table~\ref{tab:city_pm25} we report basic statistical properties of the data collected by Firefly taxis, ranking cities according to mean pollution level. Similarly, in Table~\ref{tab:EPA_pm25} we provide a ranking according to the annual mean values reported by the EPA for similar regions. While there is some agreement between the two sources of air quality measurement, differences in reporting levels of PM2.5 across regions are also apparent. 
Potential reasons for this mismatch may include differences in measurement infrastructure, including difference in sensor accuracy of reporting,
as well as differences in the way sensors are deployed geographically. In their report~\footnote{url{https://www.epa.gov/system/files/documents/2024-08/ctyfactbook2023.xlsx}} the EPA notes that \textit{the values shown are the highest among the sites in each area. Data from exceptional events are included.  This summary is not adequate in itself to numerically rank CBSAs according to their air quality.  The monitoring data represent the quality of air in the vicinity of the monitoring site and, for some pollutants, may not necessarily represent urban-wide air quality.  PM2.5 summary statistics are based on AQS data as of August 12, 2024.  All other pollutant summary statistics as of May 7, 2024}. This also explains why the EPA reported values are significantly larger to the mean PM2.5 observations recorded by the Firefly fleets.

\begin{figure}[]
  \centering
 \includegraphics[width=1.0\columnwidth]{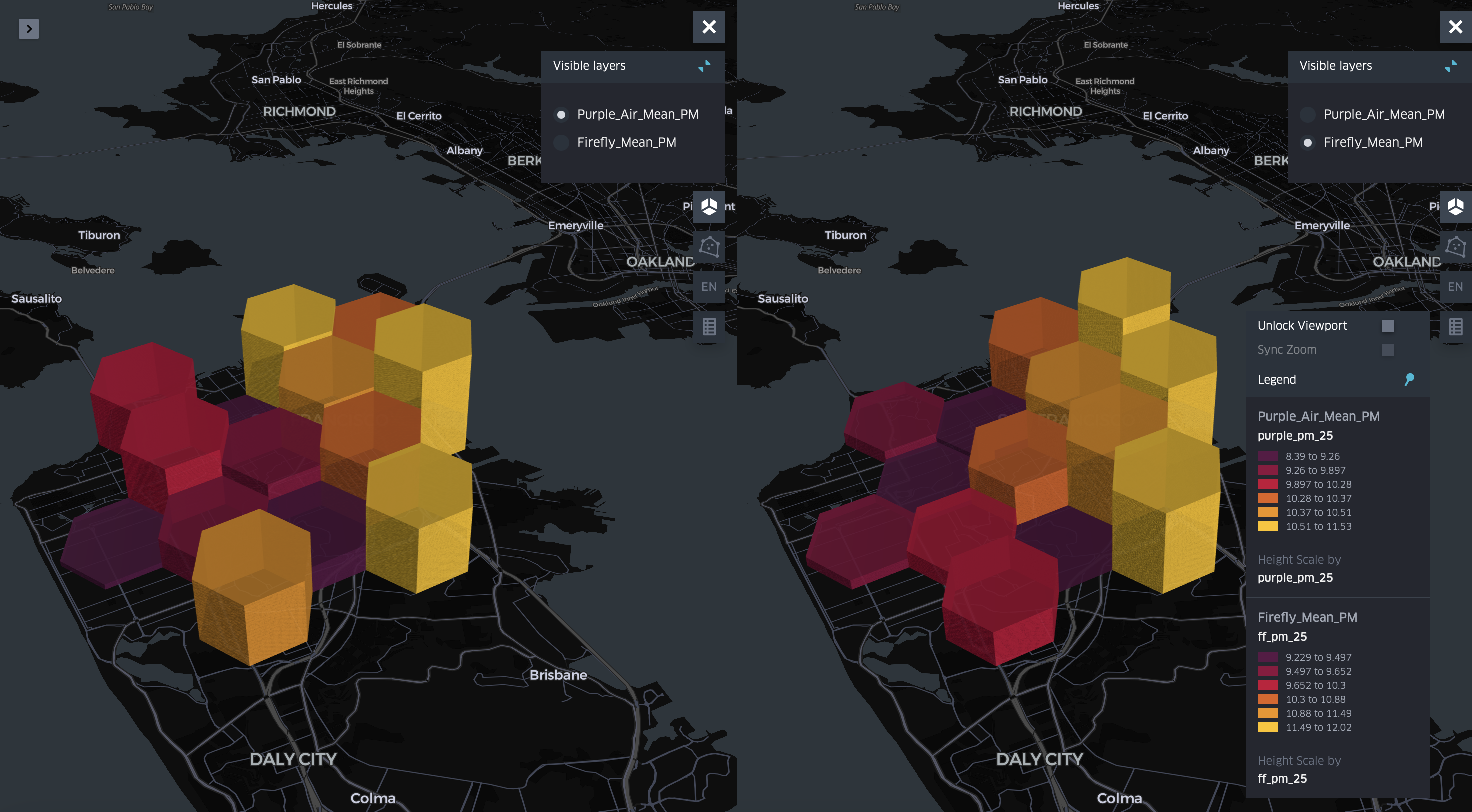}
  \caption{Geo-spatial view of mean PM2.5 levels observed for Purple Air (left) and Firefly (right) sensors.}
  \label{fig:histograms}
\end{figure}

\paragraph{Purple Air}
\label{sec:comparison}
\begin{table}[htbp]
\centering
\begin{tabular}{ccc}
\toprule
\textbf{H3\_P7 Area Index} & \textbf{Firefly Sensors} & \textbf{Purple Air Sensors} \\
\midrule
872830828ffffff & 10.91 & 10.46 \\
872830829ffffff & 9.49  & 8.39  \\
87283082affffff & 12.02 & 10.37 \\
87283082bffffff & 10.82 & 10.61 \\
87283082cffffff & 11.18 & 10.35 \\
87283082dffffff & 10.37 & 9.76  \\
87283082effffff & 11.56 & 11.53 \\
872830874ffffff & 9.61  & 10.17 \\
872830951ffffff & 10.24 & 10.37 \\
872830952ffffff & 11.56 & 10.52 \\
872830953ffffff & 9.23  & 9.24  \\
872830958ffffff & 9.51  & 8.56  \\
87283095affffff & 9.36  & 10.22 \\
87283095effffff & 9.73  & 9.36  \\
\bottomrule
\end{tabular}
\caption{Mean PM2.5 readings for different hexagonal areas within central San Francisco during September 2021. }
\label{tab:hexdata}
\end{table}
For data validation purposes, we compare the data we present here against other low cost sensors deployed in a similar region. In this regard,
we have performed a comparative analysis between Firefly sensors and Purple Air (PA) sensors, one of the most widely adopted low cost sensor projects~\footnote{\url{www2.purpleair.com}}. We consider a sample of the Firefly dataset during the initial test stages of sensor deployment trials in San Francisco from the 1st to the 30th of September 2021. We have obtained a dataset sourced through the Purple Air API that overlaps in spatio-temporal terms that allow us to compare the two data sources in alignment across the spatial and temporal dimensions. The
mean PM2.5 observations for the two datasets are visualized in Figure~\ref{fig:histograms}. The PM2.5 levels observed across the two sources are significantly close to one another even though the readings originate from sensors deployed in different locations within an area; Firefly sensors are deployed at the street-level, next to vehicular traffic, whereas Purple Air sensors tend to be deployed in buildings, typically a few floors high. 
In the measurement, we have included areas where at least three PA sensors are available during the time period of investigation and where Firefly taxis operate with high frequency. To aggregate data spatially, in both cases we use the widely adopted H3 spatial indexing system~\footnote{\url{https://h3geo.org/}} which splits the regional territory into hexagons of approximate area size equal to $5.16$ square kilometers. We provide a head to head numerical comparison of the mean PM2.5 observations for all regions visualized in Table~\ref{tab:hexdata}.

\subsection{Low cost Sensor Calibration}
Next we provide further information on the comparative analysis between taxi sensors and Purple Air (PA) sensors focusing on the application of sensor correction techniques.
The Purple Air dataset has been calibrated applying the hygroscopic correction model presented by Patel et al. in~\cite{patel2024towards}. The calibration has been based on data obtained by local reference instruments used in the US Environmental Protection Agency's Air Quality System (EPA AQS) and which measure particles under controlled, low-humidity conditions. The latter offers the opportunity to perform PM2.5 measurement that is theoretically free from humidity induced bias, which has been shown to influence measurement results according to previous studies~\cite{patel2024towards}.  
\begin{figure*}[htbp] \includegraphics[width=\textwidth]{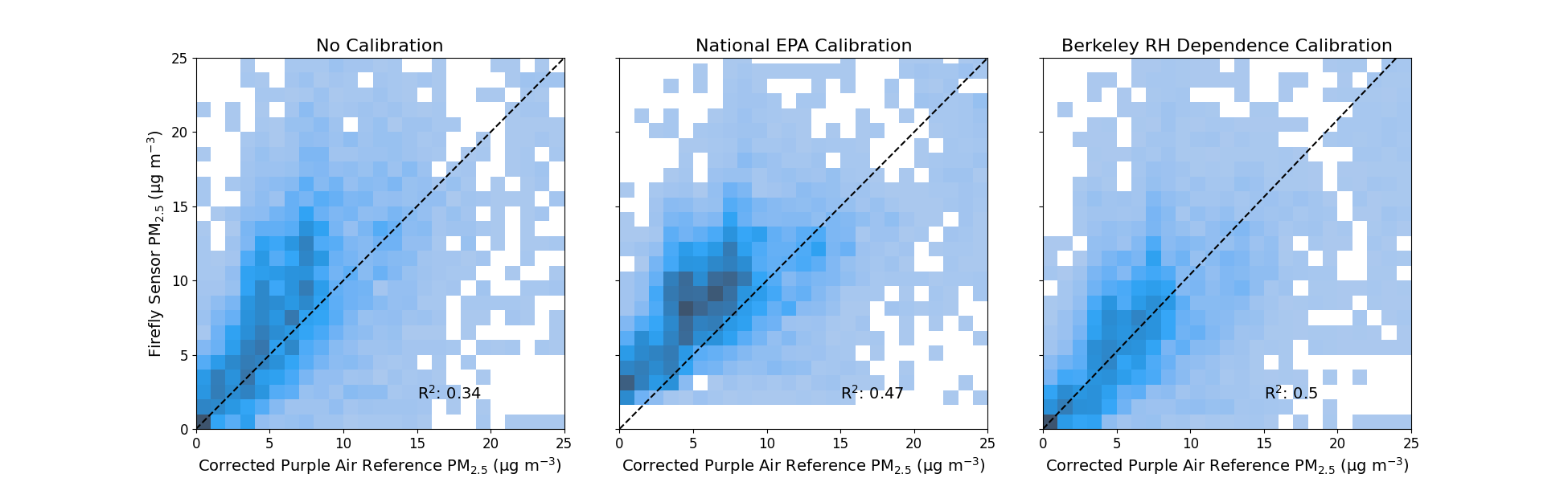}
  \caption{Firefly sensor PM2.5 values versus Purple Air reference values for September 2021 with different correction methods. The coefficient of determination (R2) is shown for each case.}
  \label{fig:calibration}
\end{figure*}



\begin{figure*}[htbp]
  \centering
  \begin{subfigure}[b]{0.48\textwidth}
    \centering
    \includegraphics[width=\textwidth]{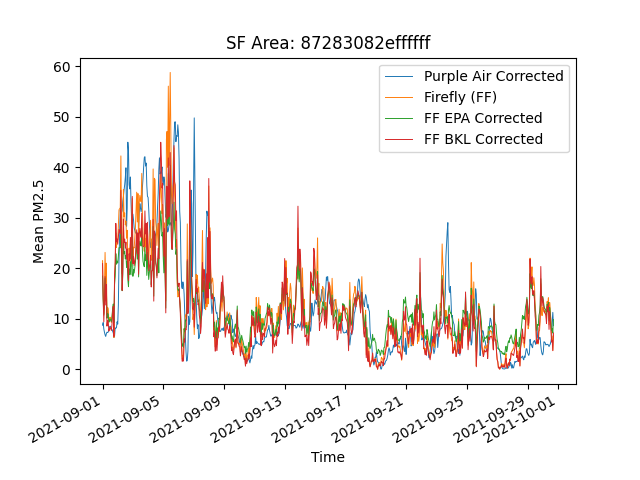}
    \caption{Time series view.}
    \label{fig:series}
  \end{subfigure}
  \hfill
  \begin{subfigure}[b]{0.48\textwidth}
    \centering
    \includegraphics[width=\textwidth]{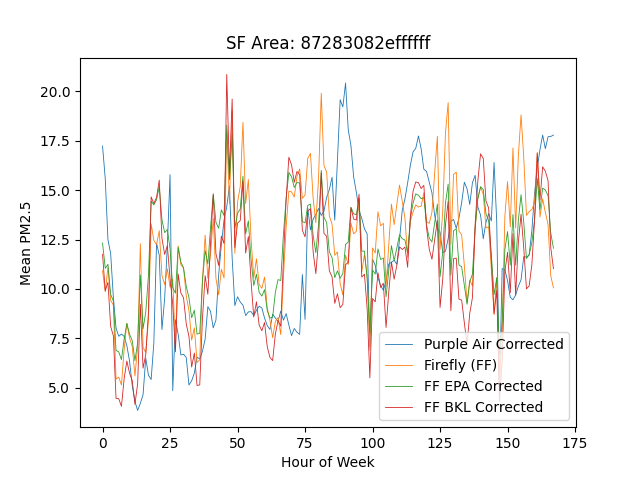}
    \caption{Weekly aggregate view.}
    \label{fig:weekly_with_humidity}
  \end{subfigure}
  \caption{Temporal views of mean PM2.5 levels aggregated at the hourly level showing signal variation for three versions of Firefly data (raw, EPA correction, Berkeley correction) and Purple Air calibrated data.}
  \label{fig:combined}
\end{figure*}

\subsubsection*{Sensor calibration and temporal views of PM2.5 readings.}
Low cost sensors, such as Purple Air devices as well as those Firefly displays are equipped with, function on the basis of enumerating PM2.5 particles according to a laser light retraction mechanism. As a consequence, particle measurement can be influenced and biased by the atmospheric makeup of aerosol particles some of which are more prone to hygroscopic growth effects. In this section, we are discussing sensor calibration techniques which aim to foster a more accurate representation of PM2.5 sensor readings. We consider the development of PM2.5 sensor calibration models an essential component of utilising low cost PM2.5 datasets such as the one presented in this work. Next, we take on the opportunity to evaluate previously proposed models~\cite{patel2024towards, barkjohn_development_2021} in a high spatio-temporal resolution setting and discuss how sensor readings compare before and after correction methods have been applied. 

Our focus is set on evaluating two sensor calibration models that incorporate relative humidity information to correct PM2.5 measurement and discuss how they compare to one another as well as raw sensor data. 
Given that there is only a single EPA sensor in the central San Francisco area~\footnote{\url{https://gispub.epa.gov/airnow/}}, we assume that sensors in closer geographic proximity to the calibration target provide a more accurate basis for correction. With this assumption in place,  we employ the signal from the nearby, corrected, Purple Air sensors as the basis to calibrate nearby moving taxi sensors. 

\subsubsection*{Calibration Methods}
The EPA has proposed~\cite{barkjohn2021development} the following linear model of PM2.5 correction using relative humidity (RH) information: 
\begin{equation}
PM_{2.5} = 0.524~\cdot~PM_{Plantower}
- 0.0862~\cdot~RH + 5.75
\end{equation}
where $RH$ corresponds to the local relative humidity levels (see here for reference to Plantower~\footnote{\url{https://amt.copernicus.org/articles/14/4617/2021/}} ). Here we consider the generalized version of this linear modeling set up, in particular 
\begin{equation}
PM_{2.5} = \alpha~\cdot~PM_{Plantower}+ \beta~\cdot~RH + \gamma
\end{equation}
The equation suggests that higher humidity levels correspond to an overestimation of PM2.5 levels as a result of hygroscopic growth ($\beta$ is smaller than $0$), which is the tendency of water particles to form connections with aerosol components. The latter has an effect on the measurement of the size distributions of microscopic particles. We also employ a sensor calibration model proposed previously in~\cite{patel2024towards}. The model calibration uses a hygroscopic growth correction factor derived from $\kappa$-Köhler theory~\cite{nilson2022development} and is employing two parameters $m$ and $k$. 

\begin{equation}
PM_{2.5} = PM_{Plantower}.\frac{m}{1+\frac{k}{100/(RH-1)}}
\end{equation}

With the introduction of the parameter $m$ the model aims to capture the distance of the assumed aerosol particle distribution at factory setting versus the actual particle distribution in the atmosphere. Parameter $k$ is introduced as means to model hygroscopic particle growth. The proposed model equation assumes the presence of non-linear interactions between particle hygroscopic growth and aerosol particle distribution differences. 

\subsubsection*{Experimental set up}
For our experimentation we employ the Firefly and Purple Air sensor datasets described in the previous paragraph and, we collect the local relative humidity levels, RH, from the PA sensors. We aggregate data for the experiment within the same hexagonal geographic areas reported above. We estimate the parameters of the EPA model through a linear optimization fit of Firefly sensors against corrected Purple air data in each area using the mean hourly observations at the area from both data sources (report parameters HERE). 

With regards to the hygroscopic growth model proposed in~\cite{patel2024towards}, we fit Firefly data to Purple air data in each area separately obtaining as a result area specific parameters k and m. Given the relatively short time window considered in the present experiment, we make the assumption that k and m are stable during the considered time period. The reader should however be cautious that these parameters may vary seasonally as described in~\cite{patel2024towards}. By fitting raw Firefly to corrected Purple air data on a rolling hourly time window during the considered month, in each area $i$, we obtain the area level $k\_{i}$ and $m\_{i}$ estimates, calculated as the mean of the respective hourly parameter estimates. We then, correct the Firefly sensor data according to these parameters. We used the code publicly available here~\footnote{\url{https://github.berkeley.edu/milan-patel/Plantower-Calibration-Paper}}. 

\subsubsection*{Calibration Results}
In Figure~\ref{fig:histograms} we present 
the raw and calibrated Firefly sensor measurements compared to the EPA Corrected Purple Air reference observations for the entire period. We observe a significant increase, beyond the mark of $30\%$, improvement on the coefficient of determination $R^{2}$ when the EPA calibration method is used, whereas a further improvement to the corrected data is seen when the $\kappa$-Köhler theory based method is applied. We highlight that while PM2.5 sensors that are deployed in the same region are expected to follow similar trends, the exact measurement at a point in time is likely to vary as sensors who are placed in different locations within the same region are likely to reflect PM2.5 density differences that may be present in the atmosphere. Taxi sensors are by deployment design closer to, and thus more likely, to measure intense levels of vehicle emissions.

In Figures~\ref{fig:series} and~\ref{fig:weekly_with_humidity} we show how Purple Air and the three versions of Firefly data compare across the temporal dimension, with Figure~\ref{fig:series} showing the time series across the period of the experiment and Figure~\ref{fig:weekly_with_humidity} showing a weekly aggregate view of the time signal of PM2.5 values. In principle, taxi sensor data demonstrates a higher degree of temporal variability suggesting perhaps that sensing takes place at a location closer to the source as vehicles in traffic emit particle forming aerosols in a manner occurring at intervals in short, sudden episodes. Purple Air sensors on the other hand show a less variable signal potentially picking PM2.5 levels at a higher atmospheric dispersion point. Despite that differences between the two sources might naturally exist, when considering the full time series we observe that sensors follow similar trends. This is also true when considering the weekly hour aggregated temporal patterns of PM2.5 (Figure~\ref{fig:weekly_with_humidity}) which suggest that despite the differences observed at very short time windows the two sensor sources detect a similar baseline pollution signal. 

\subsubsection*{Macroscopic Modeling Results for All Cities}


\begin{figure*}[htbp] 
  \centering
  \begin{subfigure}[b]{0.32\columnwidth} 
    \centering
    \includegraphics[width=\textwidth]{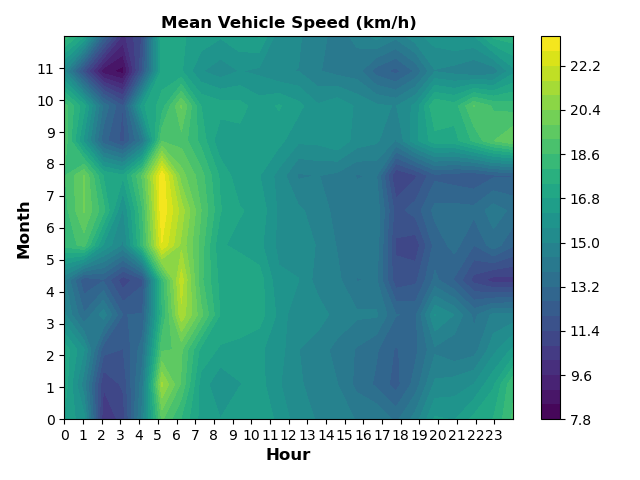}
    \caption{} 
    \label{fig:mean_speed_MIA_SI}
  \end{subfigure}
  \hfill
  \begin{subfigure}[b]{0.32\columnwidth}
    \centering
    \includegraphics[width=\textwidth]{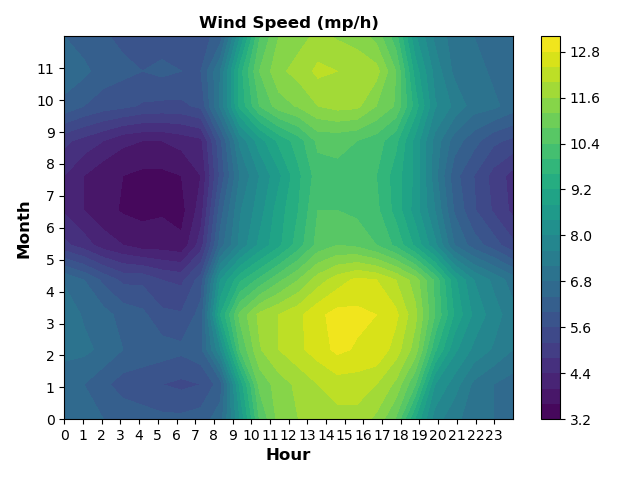}
    \caption{} 
    \label{fig:wind_speeds_MIA_SI}
  \end{subfigure}
  \hfill
  \begin{subfigure}[b]{0.32\columnwidth}
    \centering
    \includegraphics[width=\textwidth]{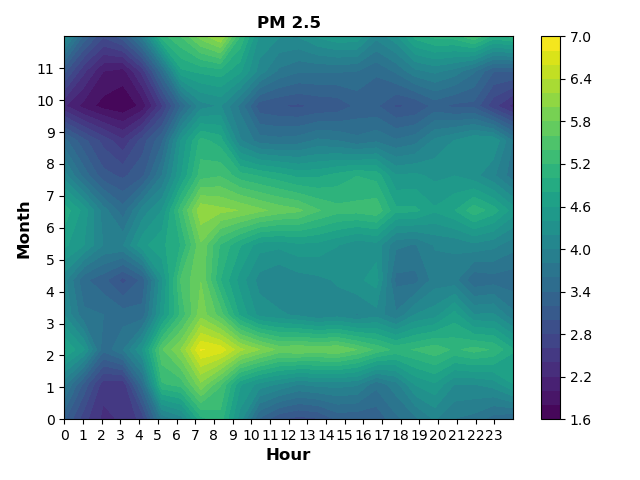}
    \caption{} 
    \label{fig:pm_MIA_SI}
  \end{subfigure}

  \caption{Supplementary contour plots representing intensity levels of mean vehicle speed (km/h), mean wind speed (km/h) and mean PM2.5 levels in the Miami area. Signal variations are shown for different months (1-12) and daily hours (0-23).}
  \label{fig:overall_MIA_SI}
\end{figure*}

\begin{figure*}[htbp]
  \centering
  \begin{subfigure}[b]{0.32\columnwidth}
    \centering
    \includegraphics[width=\textwidth]{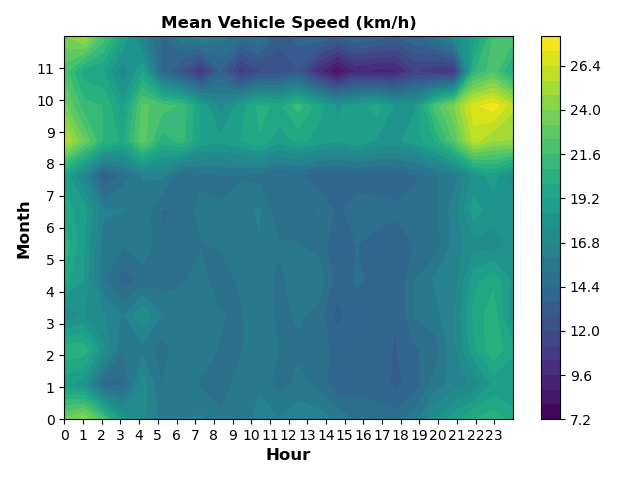}
    \caption{}
    \label{fig:mean_speed_SF_SI}
  \end{subfigure}
  \hfill
  \begin{subfigure}[b]{0.32\columnwidth}
    \centering
    \includegraphics[width=\textwidth]{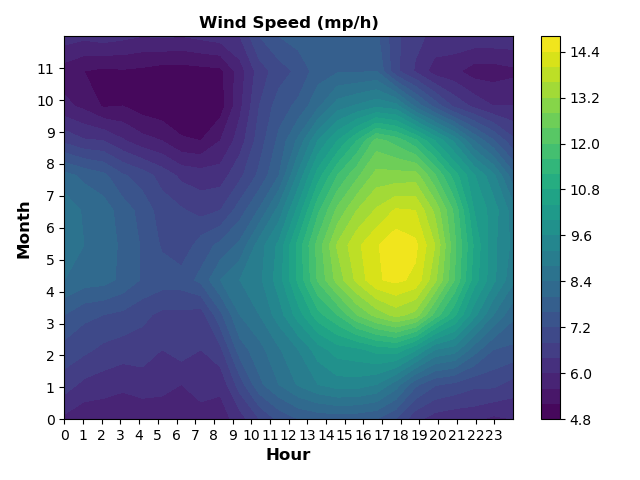}
    \caption{}
    \label{fig:wind_speeds_SF_SI}
  \end{subfigure}
  \hfill
  \begin{subfigure}[b]{0.32\columnwidth}
    \centering
    \includegraphics[width=\textwidth]{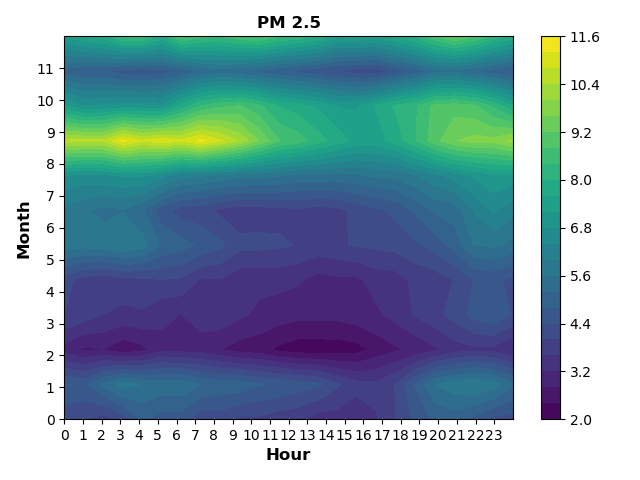}
    \caption{}
    \label{fig:pm_SF_SI}
  \end{subfigure}

  \caption{Supplementary contour plots representing intensity levels of mean vehicle speed (km/h), mean wind speed (km/h) and mean PM2.5 levels in the San Francisco area. Signal variations are shown for different months (1-12) and daily hours (0-23).}
  \label{fig:overall_SF_SI}
\end{figure*}

\begin{figure*}[htbp]
  \centering
  \begin{subfigure}[b]{0.32\columnwidth}
    \centering
    \includegraphics[width=\textwidth]{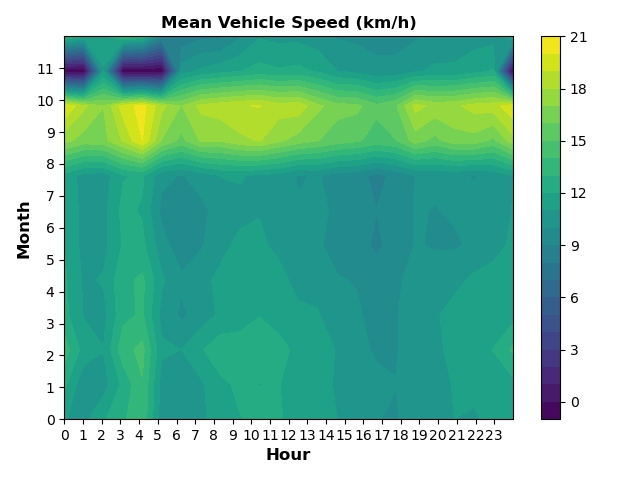}
    \caption{}
    \label{fig:mean_speed_LSV_SI}
  \end{subfigure}
  \hfill
  \begin{subfigure}[b]{0.32\columnwidth}
    \centering
    \includegraphics[width=\textwidth]{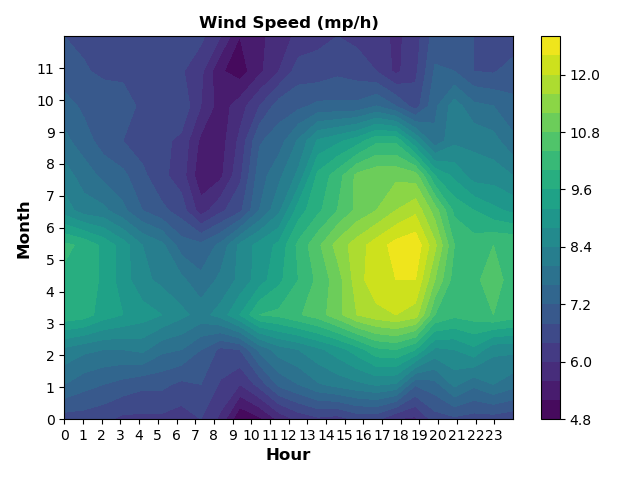}
    \caption{}
    \label{fig:wind_speeds_LSV_SI}
  \end{subfigure}
  \hfill
  \begin{subfigure}[b]{0.32\columnwidth}
    \centering
    \includegraphics[width=\textwidth]{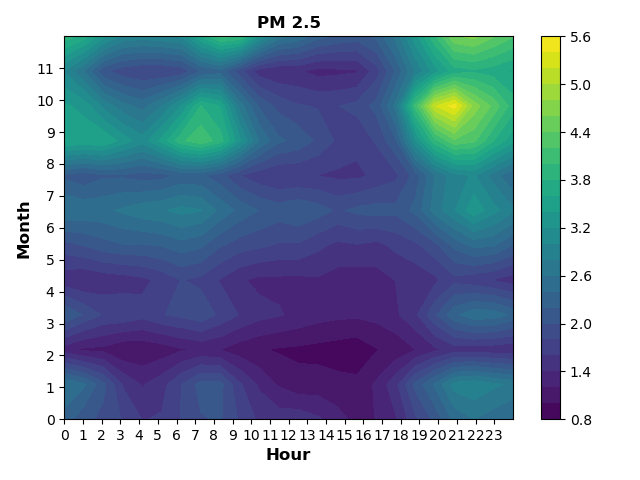}
    \caption{}
    \label{fig:pm_LSV_SI}
  \end{subfigure}

  \caption{Supplementary contour plots representing intensity levels of mean vehicle speed (km/h), mean wind speed (km/h) and mean PM2.5 levels in the Las Vegas area. Signal variations are shown for different months (1-12) and daily hours (0-23).}
  \label{fig:overall_LSV_SI}
\end{figure*}

\begin{figure*}[htbp]
  \centering
  \begin{subfigure}[b]{0.32\columnwidth}
    \centering
    \includegraphics[width=\textwidth]{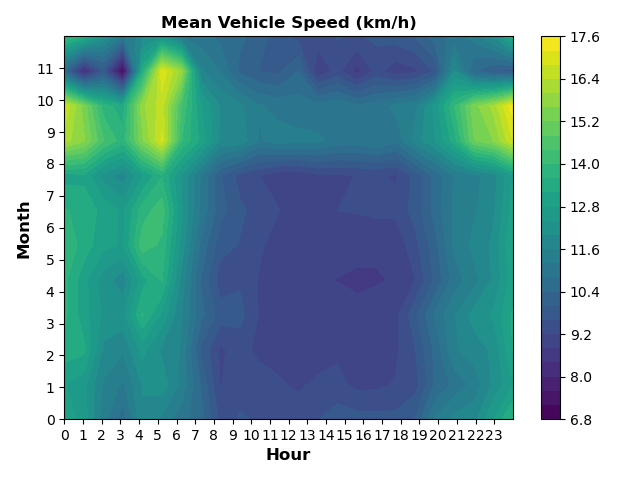}
    \caption{}
    \label{fig:mean_speed_NY_SI}
  \end{subfigure}
  \hfill
  \begin{subfigure}[b]{0.32\columnwidth}
    \centering
    \includegraphics[width=\textwidth]{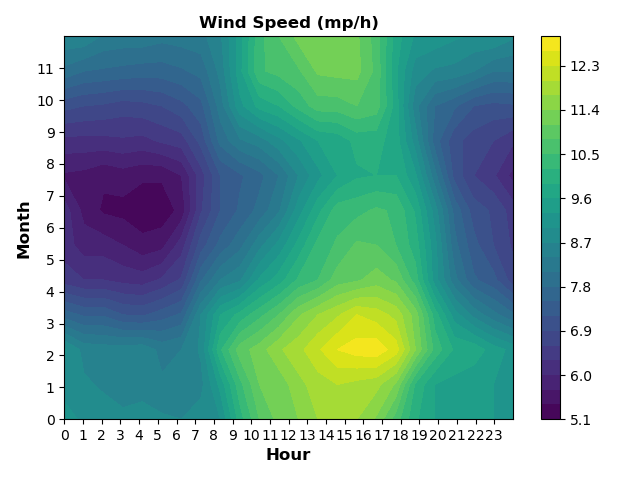}
    \caption{}
    \label{fig:wind_speeds_NY_SI}
  \end{subfigure}
  \hfill
  \begin{subfigure}[b]{0.32\columnwidth}
    \centering
    \includegraphics[width=\textwidth]{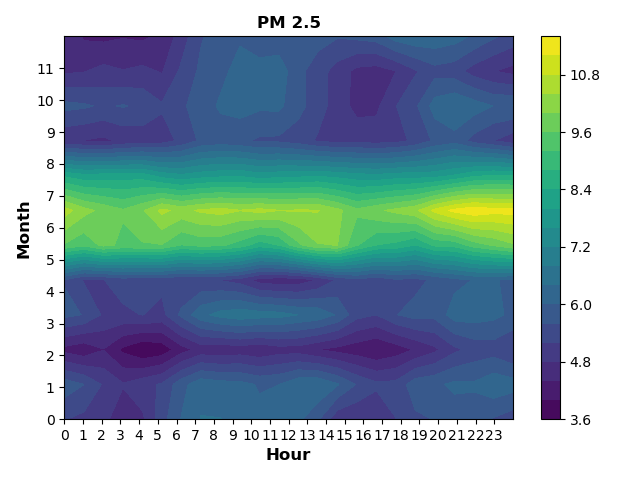}
    \caption{}
    \label{fig:pm_NY_SI}
  \end{subfigure}

  \caption{Supplementary contour plots representing intensity levels of mean vehicle speed (km/h), mean wind speed (km/h) and mean PM2.5 levels in the New York area. Signal variations are shown for different months (1-12) and daily hours (0-23).}
  \label{fig:overall_NY_SI}
\end{figure*}


\begin{figure*}[htbp]
  \centering
  \begin{subfigure}[b]{0.48\textwidth} 
    \centering
    \includegraphics[width=\textwidth]{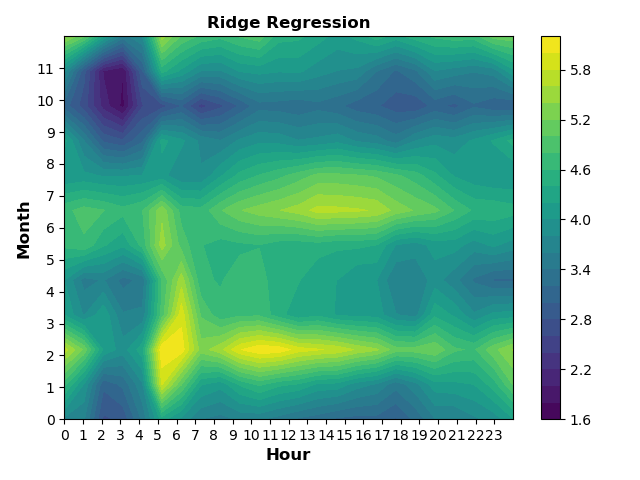}
    \caption{Miami}
    \label{fig:model_mia_monthly_SI}
  \end{subfigure}
  \hfill
  \begin{subfigure}[b]{0.48\textwidth}
    \centering
    \includegraphics[width=\textwidth]{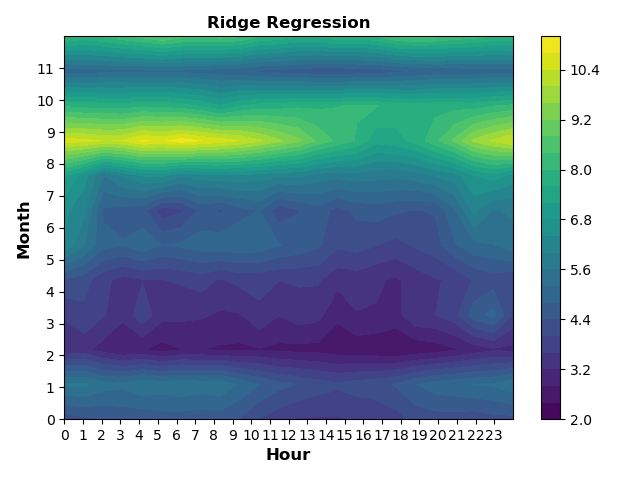}
    \caption{San Francisco}
    \label{fig:model_sf_monthly_SI}
  \end{subfigure}

  \vspace{0.5cm} 

  \begin{subfigure}[b]{0.48\textwidth}
    \centering
    \includegraphics[width=\textwidth]{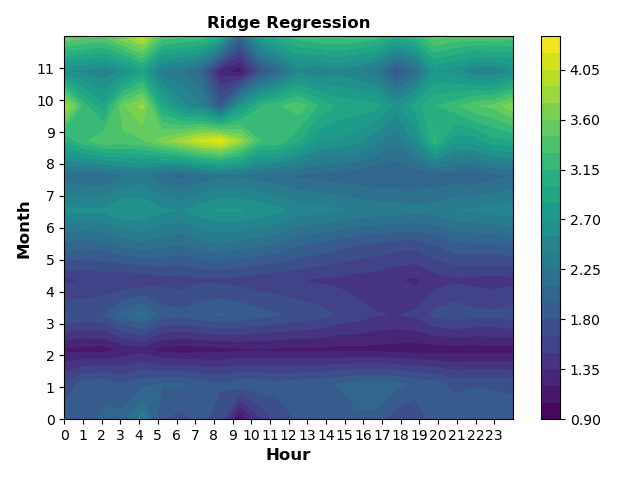}
    \caption{Las Vegas}
    \label{fig:model_lsv_monthly_SI}
  \end{subfigure}
  \hfill
  \begin{subfigure}[b]{0.48\textwidth}
    \centering
    \includegraphics[width=\textwidth]{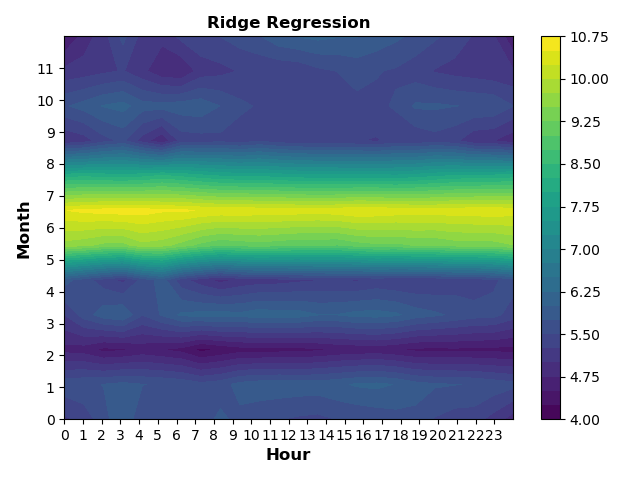}
    \caption{New York}
    \label{fig:model_ny_monthly_SI}
  \end{subfigure}

  \caption{Supplementary Congestion-Wind Regression Model Output in the Miami, San Francisco, Las Vegas, and New York Areas when considering a monthly training dataset.}
  \label{fig:modeloutputs_monthly_SI}
\end{figure*}


\begin{figure*}[htbp]
  \centering
  \begin{subfigure}[b]{0.48\textwidth}
    \centering
    \includegraphics[width=\textwidth]{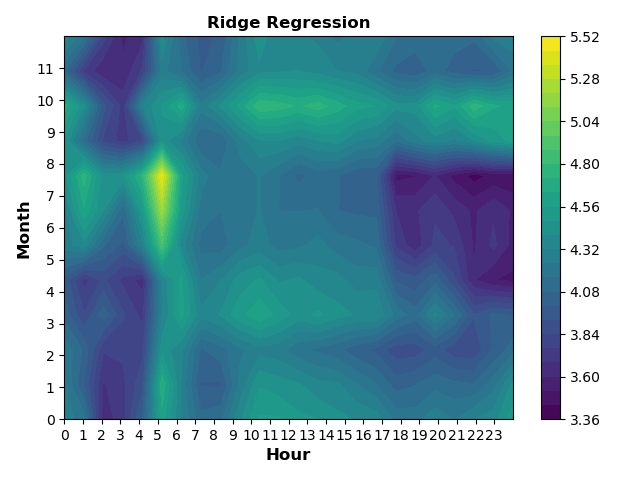}
    \caption{Miami}
    \label{fig:model_mia_annual_SI}
  \end{subfigure}
  \hfill
  \begin{subfigure}[b]{0.48\textwidth}
    \centering
    \includegraphics[width=\textwidth]{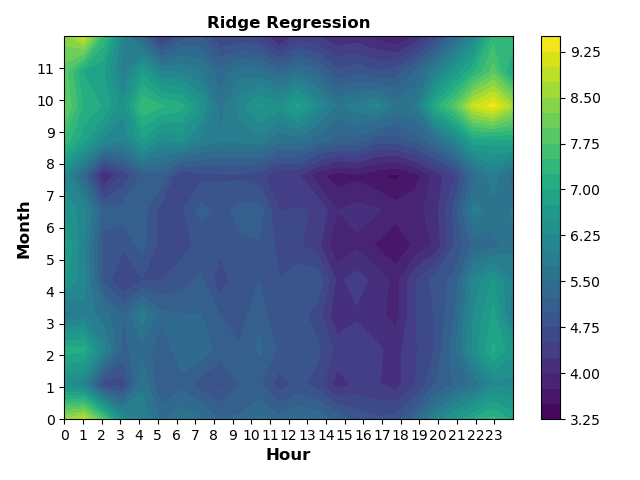}
    \caption{San Francisco}
    \label{fig:model_sf_annual_SI}
  \end{subfigure}

  \vspace{0.5cm} 

  \begin{subfigure}[b]{0.48\textwidth}
    \centering
    \includegraphics[width=\textwidth]{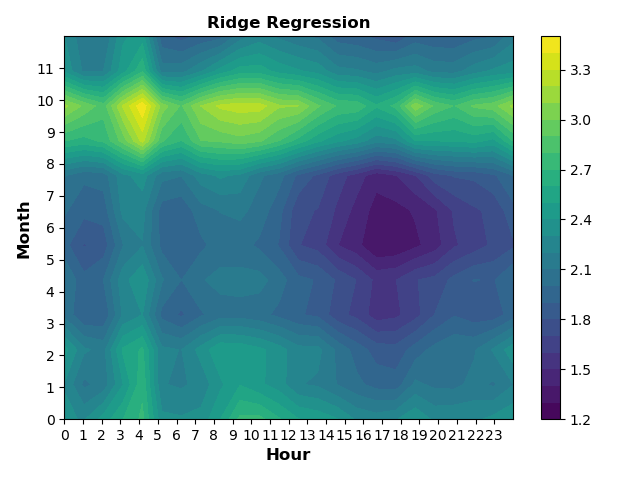}
    \caption{Las Vegas}
    \label{fig:model_lsv_annual_SI}
  \end{subfigure}
  \hfill
  \begin{subfigure}[b]{0.48\textwidth}
    \centering
    \includegraphics[width=\textwidth]{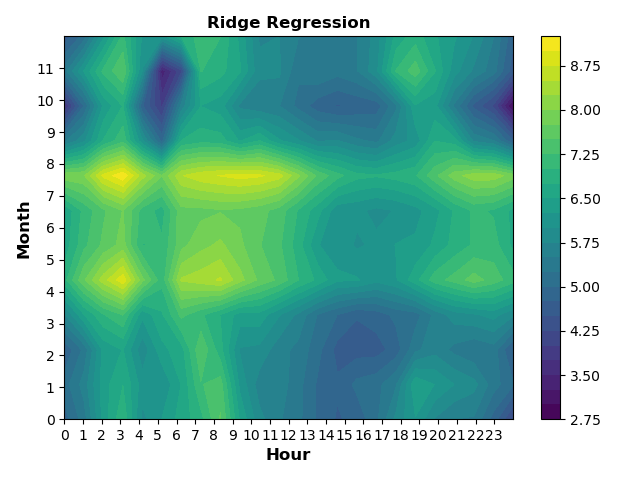}
    \caption{New York}
    \label{fig:model_ny_annual_SI}
  \end{subfigure}

  \caption{Supplementary Congestion-Wind Regression Model Output in the Miami, San Francisco, Las Vegas, and New York Areas when considering an annual training dataset.}
  \label{fig:modeloutputs_annual_SI}
\end{figure*}

\bibliographystyle{plain}
\bibliography{biblio}

\end{document}